\documentclass[iop,apj,numberedappendix,twocolappendix]{emulateapj}
\usepackage{amsmath,esint}
\usepackage{url,aasmacros}
\usepackage{natbib}
\usepackage[breaklinks,colorlinks,citecolor=blue,linkcolor=magenta]{hyperref}

\usepackage{bm}
\newcommand{\vect}[1]{\ensuremath{\boldsymbol{#1}}}

\newcommand{\snia}{{\rm SN~Ia}}
\newcommand{\sneia}{{\rm SNe~Ia}}
\newcommand{\q}[1]{{\tt #1}}

\newcommand{\Nifs}{\ensuremath{^{56}\mathrm{Ni}}}

\newcommand{\Cofs}{\ensuremath{^{56}\mathrm{Co}}}

\newcommand{\msun}{{\ensuremath{M_{\odot}}}}

\newcommand{\MCcode}{\texttt{SEDONA}}

\newcommand{\kms}{\ensuremath{\mathrm{km~s}^{-1}}}

\newcommand{\about}{\ensuremath{\sim}}

\newcommand{\dt}{\ensuremath{\Delta t}}

\newcommand{\Crfe}{\ensuremath{^{48}\mathrm{Cr}}}
\newcommand{\Vfe}{\ensuremath{^{48}\mathrm{V}}}
\newcommand{\Tife}{\ensuremath{^{48}\mathrm{Ti}}}
\newcommand{\lcdm}{\ensuremath{\Lambda}CDM}

\newcommand{\hst}{\textit{HST}}

\shorttitle{Time Delays from Chromatically Microlensed \snia\ Images}
\shortauthors{Goldstein, Nugent, Kasen, and Collett}

\begin{document}

\title{Precise Time Delays from Strongly Gravitationally Lensed Type Ia Supernovae\\with Chromatically Microlensed Images}
\author{Daniel~A.~Goldstein,\altaffilmark{1,2,$\dagger$} 
		Peter~E.~Nugent,\altaffilmark{1,2}
        Daniel~N.~Kasen,\altaffilmark{1,3,4}
        and Thomas~E.~Collett\altaffilmark{5}
}
        
\altaffiltext{1}{Department of Astronomy, University of California, Berkeley, CA 94720, USA}
\altaffiltext{2}{Computational Cosmology Center, Lawrence Berkeley National Laboratory, 1 Cyclotron Road, Berkeley, CA 94720, USA}
\altaffiltext{3}{Nuclear Science Division, Lawrence Berkeley National Laboratory, 1 Cyclotron Road, Berkeley, CA 94720, USA}
\altaffiltext{4}{Department of Physics, University of California, Berkeley, CA 94720, USA}
\altaffiltext{5}{Institute of Cosmology and Gravitation, University of Portsmouth, Dennis Sciama Building, Burnaby Road, Portsmouth, PO1 3FX, UK}

\altaffiltext{$\dagger$}{email: \href{dgold@berkeley.edu}{dgold@berkeley.edu}}

\begin{abstract}
  Time delays between the multiple images of strongly lensed Type Ia supernovae (gl\sneia) have the potential to deliver precise cosmological constraints, but the effects of microlensing on the measurement have not been studied in detail. Here we quantify the effect of microlensing on the gl\snia\ yield of the Large Synoptic Survey Telescope (LSST) and the effect of microlensing on the precision and accuracy of time delays that can be extracted from LSST gl\sneia. Microlensing has a negligible effect on the LSST gl\snia\ yield, but it can be increased by a factor of $\sim$2 to 930 systems using a novel photometric identification technique based on spectral template fitting. Crucially, the microlensing of gl\sneia\ is achromatic until 3 rest-frame weeks after the explosion, making the early-time color curves microlensing-insensitive time delay indicators. By fitting simulated flux and color observations of microlensed gl\sneia\ with their underlying, unlensed spectral templates, we forecast the distribution of absolute time delay error due to microlensing for LSST, which is unbiased at the sub-percent level and peaked at 1\% for color curve observations in the achromatic phase, while for light curve observations it is comparable to state-of-the-art mass modeling uncertainties (4\%). About 70\% of LSST gl\snia\ images should be discovered during the achromatic phase, indicating that microlensing time delay uncertainties can be minimized if prompt multicolor follow-up observations are obtained. Accounting for microlensing, the 1--2 day time delay on the recently discovered gl\snia\ iPTF16geu can be measured to 40\% precision, limiting its cosmological utility.
\end{abstract}


\section{Introduction}
\label{sec:intro}

Since the discovery of cosmic acceleration \citep{adam,saul}, \lcdm\ has become the observationally favored cosmology, implying that the universe is spatially flat, that it contains cold dark matter and baryons, and that its accelerated expansion is driven by a cosmological constant. 
Recently, a deviation from \lcdm\ was reported by \cite{riess16}, whose measurement of the Hubble constant $H_0$ using the cosmic distance ladder is in 3.4$\sigma$ tension with the value inferred from the cosmic microwave background \citep[CMB;][]{planck15}, assuming a $\Lambda$CDM cosmology and the standard model of particle physics.
Independent measurements of $H_0$ with percent-level accuracy are necessary to determine whether the discrepancy is due to new physics (e.g., a new neutrino species; \citealt{riess16,bonvin17}) or to systematics.

Strong gravitational lensing is an independent probe of the cosmological parameters \citep{oguri12,suyu13,collett14}. 
Time delays between multiple images of strongly gravitationally lensed variable sources are particularly sensitive to $H_0$, making them ideal tools to test this discrepancy.
They also are sensitive to other parameters of the cosmological model, such as the dark energy equation of state and its evolution with redshift \citep{linder04,linder11,cosmography}.

\begin{figure*}
	\centering
    \includegraphics[width=1\textwidth]{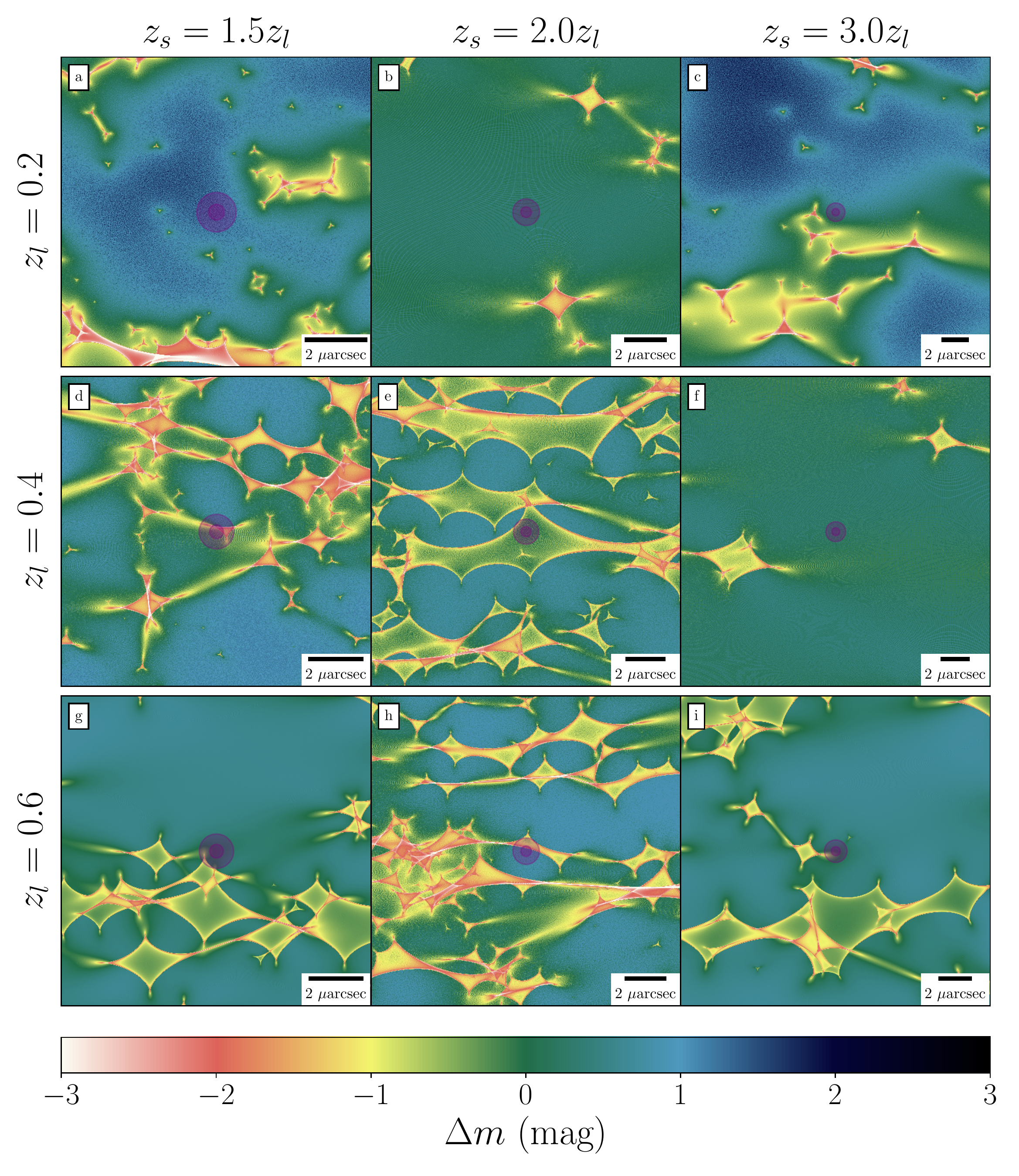}
    \caption{Source-plane magnification patterns of nine of 78,184 the lens galaxy star fields considered in this analysis.
    Each panel consists of 1,000$^2$ pixels and has a side length of 10 times the Einstein radius of a 1\msun\ deflector projected onto the source plane.
    The detailed parameters of each map are given in Table \ref{tab:mmpars}.
    The size of the exterior shell ($4 \times 10^4$ \kms) of the \snia\ model W7 at 20 (50) days after explosion is plotted as the interior (exterior) purple circle at the center of each map.
	Negative (positive) $\Delta m$ indicates magnification (demagnification) over the value from a smooth mass model without microlensing.
}
    \label{fig:mmap}
\end{figure*}

Measuring cosmological parameters to percent level accuracy with strong lens time delays requires three main ingredients \citep{suyu17}. 
First, one must measure the time delays \citep[e.g.][]{tewes13,bonvin17}. 
Second, the lensing potential must be inferred to convert the observed time delays into measurements of the time delay distance \citep[e.g.,][]{wong17}. 
This relies on reconstruction of the extended features of a lensed host. 
Finally, the effect of weak lensing by mass close to the lens and along the line of sight must be included \citep[e.g.,][]{suyu10,collett13,rusu17,mccully17}, since lenses are typically found in overdense regions of the universe \citep{fassnacht11}.

To date, time delay cosmography has only been attempted with strongly lensed active galactic nuclei \citep[AGNs; e.g.,][]{vuissoz08,suyu13,tewes13,bonvin16}. 
Lensed AGNs complicate these ingredients, making percent-level constraints on $H_0$ difficult. 
Because the light curves of AGNs are stochastic and heterogeneous, they typically require years of cadenced monitoring to yield precise time delays \citep{tdc}. 
Inferring the lensing potential by reconstructing the lensed host light is challenging since AGNs typically outshine their host galaxies by several magnitudes.
Detecting lensed AGNs requires observations of multiple images introducing a selection function for larger Einstein radii and hence an overdense line of sight \citep{collettcunnington16}, leading to systematic overestimates of $H_0$.

In contrast, the light curves of Type Ia supernovae (\sneia) are remarkably homogeneous, and strongly lensed \snia\ (gl\snia) light curves evolve over weeks, not years, allowing their time delays to be measured with far less observational overhead than those of AGNs. 
In addition, gl\sneia\ fade away, allowing a simpler reconstruction of the lensed hosts. 
gl\sneia\ can be detected without resolving multiple images \citep{goobar16}, simplifying the selection function.
Because gl\sneia\ are standardizable candles, they also have the potential to directly determine the lensing magnification factor $\mu$, which breaks the degeneracy between the lens potential and the Hubble constant \citep{oguri03}, if the microlensing and macrolensing magnifications can be separated.
The well known spectral energy distributions (SEDs) of \sneia\ also allow one to correct for extinction along the paths of each \snia\ image---another major advantage over AGNs.

So far, only one gl\snia, iPTF16geu, has been discovered with resolved images \citep{goobar16}. 
However, future surveys, especially the Large Synoptic Survey Telescope \citep[LSST;][]{lsst}, are expected to discover hundreds \citep{gn17}.
Thus the prospects for discovering a sufficient number of gl\sneia\ to perform time-delay cosmography in the near future are good.

However, there is a foreground that threatens this outlook: microlensing. 
It has long been known that lens galaxy field stars can significantly magnify and demagnify cosmologically distant background AGNs \citep{ml} and supernovae \citep{dk06,bag06}.
In the microlensing scenario, different macroimages of the same source propagate through different regions of the lens galaxy, passing through distinct lens galaxy star fields. 
The star fields possess rich networks of caustics that introduce magnification patterns into the source plane.
These patterns vary over characteristic angular scales of microarcseconds (hence ``microlensing," see Figure \ref{fig:mmap}), which are typically comparable to the physical sizes of supernovae and AGNs.
Thus, as a strongly lensed supernova expands over the source plane, it can experience time- and wavelength-dependent magnifications unique to each lensed image, distorting their light curves and spectra in different ways. 
These distortions can make it harder to ``match up'' the light curves of multiple images and extract an accurate time delay.\footnote{N.B. The general relativistic time delays introduced by microlensing into macroimages are of order microseconds \citep{moore96}---too small to detect.
The uncertainty microlensing introduces into time delays is solely due to time- and wavelength-dependent distortions of macroimage light curves and spectra.}

Microlensing of lensed variable sources is more than a theoretical exercise: it has been reported in many strongly lensed quasars \citep[e.g.,][]{cml}, and there is also evidence that it affects the images of iPTF16geu \citep{more16}.
\cite{dk06} estimated that microlensing can introduce uncertainties of several days into the features of supernova light curves that can yield time delays.
Typical time delays for gl\sneia\ are a couple of weeks \citep{gn17},  translating to a typical fractional time delay uncertainty of (a few days) / (two weeks) $\sim20\%$. 
At this precision, $\sim$400 gl\snia\ time delays would be required to reach a 1\% uncertainty on $H_0$, assuming no other sources of error, whereas a single gl\snia\ time delay with 1\% precision could accomplish the same goal.
Thus, controlling microlensing is of critical importance to the success of time delay cosmography with gl\sneia.

\begin{deluxetable}{lccccc}
\tablecaption{Properties of the magnification patterns in Figure \ref{fig:mmap}.}
\tablehead{& $\kappa$ & $f_*$ & $\gamma$ & $q$ & $\langle \mu \rangle$ }
\startdata
a & 1.30 & 0.84 & 1.30 & 0.20 & 0.63\\
b & 0.24 & 0.35 & 0.25 & 0.10 & 1.93\\
c & 1.25 & 0.83 & 1.26 & 0.20 & 0.65\\
d & 0.87 & 0.72 & 0.93 & 1.00 & 1.19\\
e & 0.75 & 0.80 & 0.76 & 1.00 & 1.96\\
f & 0.30 & 0.38 & 0.27 & 1.00 & 2.39\\
g & 0.36 & 0.49 & 0.33 & 1.00 & 3.30\\
h & 0.84 & 0.72 & 0.79 & 0.10 & 1.66\\
i & 0.37 & 0.50 & 0.31 & 1.00 & 3.29
\enddata
\tablecomments{$\kappa$: local convergence. 
$f_*$: fraction of surface density in stars. 
$\gamma$: local shear. 
$q$: mass ratio $m_{min}/m_{max}$ of the stellar mass function.
$\langle \mu \rangle$: magnification of the field from a smooth mass model without microlensing.}
\label{tab:mmpars}
\end{deluxetable}

If, as proposed by \cite{gn17}, the microlensing magnification affects all wavelengths equally (i.e., if it is ``achromatic'')---then one could use the color curves of gl\sneia\ instead of the broadband light curves to extract time delays even if the images are affected by microlensing. 
For example, for a given image, if the $B$-band is macro- and microlensed as much as the $U$-band, then in the $U-B$ color curve, the micro- and macrolensing magnifications will both cancel out, leaving features common to the color curves of all images that can  pin down the time delays to high precision.
This would enable color curves of different images to be compared meaningfully, yielding time delays with less uncertainty.

In this article, we use detailed radiation transport simulations of a well-understood \snia\ model to assess the viability of extracting time delays from the color curves of gl\sneia\ to circumvent the effects of microlensing.
We also perform the first gl\snia\ yield calculation that takes microlensing into account.
The structure of the paper is as follows.
In Section \ref{sec:sim}, we describe the radiation transport, gl\snia\ population, and microlensing models used to synthesize representative microlensed  gl\snia\ SEDs.
In Section \ref{sec:chrom}, we present the results of our simulations and use them to show that the microlensing of gl\sneia\ exhibits an achromatic phase at early times.
In Section \ref{sec:salt}, we present a novel spectral template-based gl\snia\ photometric detection technique and use it to forecast the gl\snia\ yield of LSST.
In Section \ref{sec:dtmeth}, we forecast the time delay uncertainty due to microlensing and show that it can be controlled to 1\% for typical LSST systems. 
We conclude in Section \ref{sec:conclusion}.
Throughout this paper we assume a \cite{planck15} cosmology with $\Omega_\Lambda=0.6925$, $\Omega_m=0.3075$, and $h=0.6774$.

\section{Population, Radiation Transport, and Microlensing Simulations}
\label{sec:sim}
In this section, we describe the simulation framework we use to generate a realistic population of gl\sneia.
First, the framework realizes a population of unlensed \sneia\ and elliptical galaxy lenses using measured redshift distributions. 
It solves the lens equation for supernovae and lenses close together on the sky, and when a multiply imaged system is produced, it yields image multiplicities, time delays, magnifications, and image positions.
For each lensed supernova image, the framework generates a microlensing magnification pattern based on the image properties. 
A theoretical \snia\ spectral time series is convolved with the magnification pattern, giving the lensing amplification of the supernova SED as a function of time and wavelength. 
This in turn is applied to an empirical \snia\ SED template.  
Realistic LSST light curves are generated from these microlensed spectral templates and fed to a novel detection algorithm.

\subsection{The Strongly Lensed Type Ia Supernova Population}
\label{sec:population}
In the present analysis we use the same gl\snia\ population model as \cite{gn17}.
We consider only elliptical galaxy lenses and model their mass distribution as a Singular Isothermal Ellipsoid \cite[SIE;][]{kormann94}, which has shown excellent agreement with observations \citep[e.g.,][]{koopmans09}.
The SIE convergence $\kappa$ is given by:
\begin{equation}
\label{eq:sie}
\kappa(x,y) = \frac{\theta_{E}}{2}
	\frac{\lambda(e)}{\sqrt{(1-e)^{-1}x^2+(1-e)y^2}},
\end{equation}
where
\begin{equation}
\label{eq:einrad}
\theta_{E} = 4 \pi \left(\frac{\sigma}{c}\right)^2\frac{D_{ls}}{D_s}.
\end{equation} 
In the above equations, $\sigma$ is the velocity dispersion of the lens galaxy, $e$ is its ellipticity, and $\lambda(e)$ is its so-called ``dynamical normalization,'' a parameter related to three-dimensional shape.
Here we make the simplifying assumption that there are an equal number of oblate and prolate galaxies, which \cite{chae03} showed implies $\lambda(e) \simeq 1$. 
As in \cite{oguri08}, we assume $e$ follows a truncated normal distribution on the interval $[0.0, 0.9]$, with $\mu_e= 0.3$, $\sigma_e = 0.16$. 

We also include external shear to account for the effect of the lens environment \citep[e.g.,][]{kochanek91, keeton97, witt97}
We assume $\log_{10}\gamma_\mathrm{ext}$ follows a normal distribution with mean $-1.30$ and scale 0.2, consistent with the level of external shear expected from ray tracing in $N$-body simulations \citep{holder03}.
The orientation of the external shear is assumed to be random.

We model the velocity distribution of elliptical galaxies as a modified Schechter function \citep{sheth03}:
\begin{equation}
\label{eq:schechter}
dn = \phi(\sigma) d\sigma = \phi_*\left(\frac{\sigma}{\sigma_*}\right)^\alpha \exp\left[-\left(\frac{\sigma}{\sigma_*}\right)^\beta\right]\frac{\beta}{\Gamma(\alpha/\beta)}\frac{d\sigma}{\sigma},
\end{equation}
where $\Gamma$ is the gamma function, and $dn$ is the differential number of galaxies per unit velocity dispersion per unit comoving volume.
We use the parameter values of \cite{choi07} from the Sloan Digital Sky Survey \citep[SDSS;][]{sdss}:
$(\phi_*, \sigma_*, \alpha, \beta) = (8 \times 10^{-3}~h^3~\mathrm{Mpc}, 161~\kms,2.32, 2.67)$.
We assume the mass distribution and velocity function do not evolve with redshift, consistent with the results of \cite{chae07}, \cite{oguri08}, and \cite{bezanson11}. 

To convert Equation \ref{eq:schechter} into a redshift distribution, we use the definition of the comoving volume element:
\begin{equation}
	dV_C = D_H \frac{(1+z)^2 D_A^2}{E(z)}~dzd\Omega,
\end{equation}
where $D_H = c / H_0$ is the Hubble distance, $E(z) = \sqrt{\Omega_M(1+z)^3 + \Omega_\Lambda}$ in our assumed cosmology, and $D_A$ is the angular diameter distance.
Since $dn = dN/dV_C$, for the unnormalized all-sky $(d\Omega = 4\pi)$ galaxy distribution we have
\begin{equation}
\label{eq:galaxydist}
\frac{dN}{d\sigma dz} =  4\pi D_H \frac{(1+z)^2 D_A^2}{E(z)}\phi(\sigma).
\end{equation}

Integrating Equation \ref{eq:schechter} over $0 \leq z \leq 1$ and $10^{1.7}~\kms \leq \sigma \leq 10^{2.6}~\kms$, we find that there are $N_\mathrm{gal} \simeq 3.8 \times 10^8$ elliptical galaxies, all sky, that can act as strong lenses. 
This gives the joint probability density  function for $\sigma$ and $z$:
\begin{equation}
\label{eq:galaxypdf}
p(\sigma, z) = \frac{1}{N_\mathrm{gal}} \frac{dN}{d\sigma dz}.
\end{equation}

\sneia\ exhibit a redshift-dependent volumetric rate and an intrinsic dispersion in rest-frame $M_B$. 
In our model of the \snia\ population, we take the redshift-dependent \snia\ rate from \cite{2000MNRAS.319..549S}.
We assume that the peak rest-frame $M_B$ is normally distributed with $\mu_M = -19.3$ and $\sigma_M = 0.2$.
For simplicity, we neglect extinction.

The lens and source populations are realized in a Monte Carlo simulation.
We generate $10^5$ lens galaxies with parameters drawn at random from their underlying distributions.
For each lens galaxy, an effective lensing area of influence is estimated as a $[8 \theta_{E, z_s=\infty}]^2$ box centered on the galaxy.\footnote{This box size was chosen to be large enough to accommodate the effects of ellipticity and external shear.}
We simulate $5 \times 10^4$ years of \sneia, randomly distributed across the box, rejecting systems where $z_s < z_l$.
For each remaining source, we solve the lens equation using  \q{glafic}\  \citep{glafic} to determine the macrolensing magnification, image multiplicity, and time delays.
In total we generated 37,100 multiply imaged systems containing a total of 78,184 images.
Since our simulation only covers $10^5 / N_\mathrm{gal} \approx 0.026\%$ of the sky, this  corresponds to a rate of 2,675 systems, all sky, per year to $z_s=2$. 

\subsection{Microlensing Magnification Patterns}
\label{sec:ml}

For each image we calculate a source-plane magnification pattern with \texttt{microlens} \citep{wthesis,mltree}, an inverse ray-tracing code.
In this scheme, stars modeled as point-mass deflectors are realized from a mass function at random locations in a two-dimensional field of the lens galaxy.
The size of the pattern is characterized by the Einstein radius $\bar{R}_E$ of a deflector of mass $\bar{m}$ projected onto the source plane,
\begin{equation}
\bar{R}_E = \sqrt{\frac{4G\bar{m}}{c^2}\frac{D_{ls}D_s}{D_l}},
\end{equation}
where $D_l$ is the angular diameter distance to the lens, $D_s$ is the angular diameter distance to the source, and $D_{ls}$ is the angular diameter distance between the lens and the source.
Our magnification patterns are 10$\bar{R}_E$ on a side, which for typical source and lens redshifts ($z_s = 1.2, z_l = 0.6$) corresponds to an angular scale of $10\bar{R}_E / D_s \approx 1.5 \times 10^{-5}$ arcsec.
At the same redshifts, $\bar{R}_E \approx 2.7 \times 10^3$ AU, which is roughly 5 times larger than the extent of the supernova model near peak brightness.

The magnification patterns are specified by four parameters:
(1) the local convergence, $\kappa$, 
(2) the local shear, $\gamma$, including the contributions of both the SIE and external potentials,
(3) the fraction of the local convergence in stars, $f_*$,\footnote{The remainder of the convergence is assumed to take the form of continuously distributed matter (i.e., dark matter).} and
(4) the dynamic range of the stellar mass function, $q=m_{min}/m_{max}$.
Supplying these parameters allows \texttt{microlens}\ to solve the general microlensing equation, $\vect{\beta} = \vect{\theta} - \vect{\alpha}$, which resolves to
\begin{equation}
\label{eq:lenseq}
\vect{\beta} =  \begin{pmatrix}
1 - \gamma - \kappa_c & 0 \\
0 & 1 + \gamma - \kappa_c
\end{pmatrix}\vect{\theta} - \sum_{i=1}^{N_*}\frac{M_i (\vect{\theta} - \vect{\theta}_i)}{(\vect{\theta} - \vect{\theta}_i)^2},
\end{equation}
where $\kappa_c=(1-f_*)\kappa$ is the local convergence in continuously distributed matter, the two-dimensional vector $\vect{\beta}$ is the angular position of the source in the absence of lensing, $\vect{\theta}$ is the angular position of the observed macroimage, $\vect{\theta}_i$ is the angular position of the $i$'th star, and $N_*$ is the number of stars in the  field, determined from the local convergence in stars $\kappa_* = f_*\kappa$ using the procedure of \cite{schneiderweiss87}.

\begin{figure*}
	\centering
    \includegraphics[width=1\textwidth]{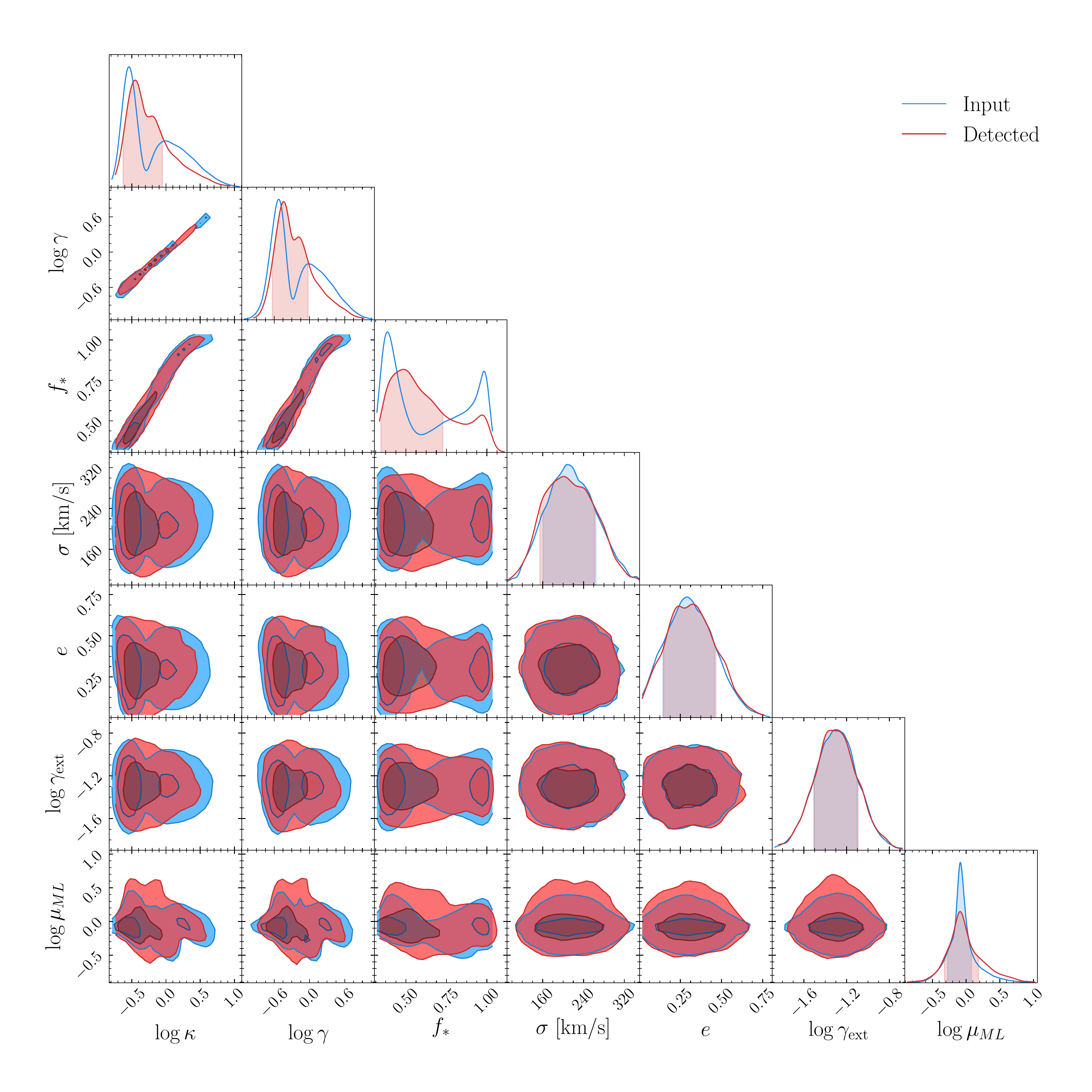}
    \caption{Two-dimensional projections of the joint distributions of lensed image macrolensing and microlensing parameters.
    The input distributions (blue) represent all 78,184 simulated lensed images from Section \ref{sec:population}, and the detected distributions (red) represent only the images of the gl\sneia\ detected in Section \ref{sec:detectmc}.
    The bimodal input distributions of $f_*$, $\gamma$, and $\kappa$ represent the amplified $(f_* \approx 0.4)$ and overfocussed $(f_* \approx 1)$ images produced by SIE lenses. 
    Joint contours show 1 and 2$\sigma$. 
    Marginal shaded regions show 1$\sigma$.
    }
	\label{fig:mlsl}
\end{figure*}

We use a \cite{salpeter} mass function, $dn/dm \propto m^{-2.35}$, to model the population of stars in our microlensing calculations.
As we will show in Section \ref{sec:chrom}, this choice has no effect on our results as the achromaticity of gl\snia\ microlensing is driven entirely by the color evolution of \sneia\ and not by the properties of the microlensing magnification patterns. 
Following \cite{dk06}, we take the mean mass $\bar{m} = 1 \msun$.
The microlensing parameters $\kappa$ and $\gamma$ are determined by evaluating the SIE and external shear lensing potentials at the  location of each image.
$f_*$ is estimated following the method of \cite{dk06}, assuming a de Vaucouleurs stellar profile normalized so that the maximum $f_* = 1$.
For each image, the dynamic range parameter $q$ is sampled uniformly at random from $(0.1, 0.2, 0.5, 1.0)$, appropriate for the old stellar populations in elliptical galaxies.
Figure \ref{fig:mmap} shows a random selection of nine of the maps, highlighting their morphological diversity. 
Figure \ref{fig:mlsl} shows two dimensional projections of the joint distributions of the macrolensing parameters $\sigma$, $e$, and $\gamma_\mathrm{ext}$, the microlensing parameters $\kappa$, $\gamma$, and $f_*$, and the time- and wavelength-averaged microlensing magnification $\mu_{ML}$. 

\subsection{Supernova Modeling}
\label{sec:sn}
We use the well-understood, spherically symmetric\footnote{Spectropolarimetry indicates that \sneia\ are globally spherically symmetric to $\sim$a few percent. See \cite{specpol}.} \snia\ atmosphere model W7 \citep{w7} to estimate the time- and wavelength-dependent magnification of gl\sneia\ due to microlensing.
This radiation transport model is the result of a one-dimensional explosion simulation in which a Chandrasekhar-mass carbon-oxygen white dwarf undergoes a deflagration.
The explosion of the white dwarf completely unbinds the star and deposits the energy liberated by nuclear burning into the ejected mass. 
The deposited energy controls the velocity distribution of the ejecta and its density profile, which is assumed to reach homology seconds after the explosion. 
We use the time-dependent, Monte Carlo radiation transport code \MCcode\ \citep{sedona} to calculate the spectral time series of the model.
Details of our \MCcode\ simulations appear in Appendix \ref{sec:sedona}.

The observed spectrum $F_\lambda$ of the model at wavelength $\lambda$ and time $t$ is obtained by convolving its time-evolving specific intensity with the lensing amplification pattern over a plane normal to the observer's line of sight.
Since the model is spherical, this integral takes the form:
\begin{equation}
	\label{eq:lspec}
	F_\lambda(\lambda, t) = D_L^{-2} \int_0^{2\pi}\,\int_0^{P_m} I_\lambda(P, \phi, \lambda, t)\, \mu(P, \phi)\,P\,dP\,d\phi,
\end{equation}
where $\phi$ and $P$ are azimuthal and impact parameter coordinates on the plane, $I_\lambda$ is the specific intensity of the model, $\mu$ is the lensing amplification,\footnote{In this paper, $\mu$ refers exclusively to lensing amplification. 
Nowhere should $\mu$ be interpreted as $\mu = \cos \theta$, the viewing angle parameter that frequently appears in supernova modeling papers.} $D_L$ is the luminosity distance to the supernova, and $P_m$ is the maximum impact parameter of the model.
For a derivation of Equation \ref{eq:lspec}, see Appendix \ref{sec:geom}.
The time- and wavelength-dependent magnification of a given magnification pattern is obtained by dividing $F_\lambda$ by the unlensed spectrum of the model (Equation \ref{eq:lspec} with $\mu=1$).

We interpolate each magnification pattern bilinearly and convolve it with the redshifted\footnote{We refer here to cosmological redshift only; Doppler shifts due to supernova expansion velocity are accounted for implicitly in the radiation transport simulation.} specific intensities of the supernova model.
The redshift configurations control the projected size of the supernova on the magnification pattern and thus the magnification experienced by each differential element of the projected supernova atmosphere.
They also control $\theta_E$ and thus $\kappa$, $\gamma$, and $f_*$ at the location of the image.
We always place the supernova model at the center of the magnification pattern. 
We model the homologous expansion of the supernova behind the magnification pattern (i.e., the projected size of the supernova on the magnification pattern changes with time), but not relative motion between the supernova and the lens galaxy star field. 
In general, supernova atmospheres both expand and move with respect to the lens galaxy, but the characteristic expansion velocity of the atmosphere ($\sim$10$^4\, \mathrm{km}\,\mathrm{s}^{-1}$) is much larger than the characteristic relative velocity between the lens galaxy and the supernova ($\sim$10$^2\,\mathrm{km}\,\mathrm{s}^{-1}$), so here we model only the effects of expansion.

\section{Two Phases of Type Ia Supernova Microlensing}
\label{sec:chrom}
\begin{figure*}
	\centering
    \includegraphics[width=1\textwidth]{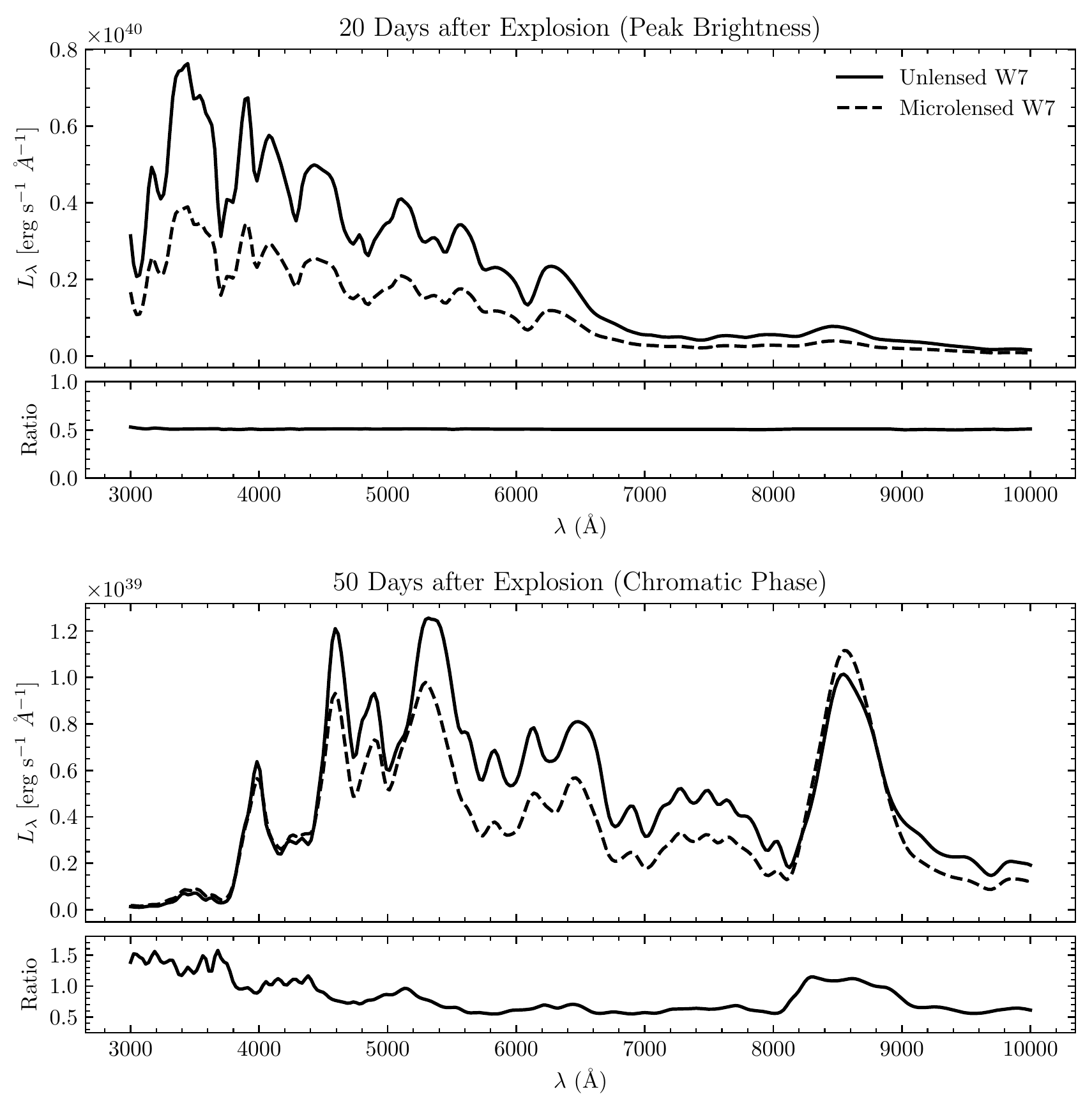}
    \caption{Rest-frame spectra of model W7 \citep{w7} computed with \MCcode\ near peak brightness and well into the onset of the infrared secondary maximum.
    The solid curve is unlensed and the dashed curve is lensed by star field (h) from Figure \ref{fig:mmap}. 
    Near peak brightness, microlensing does not have a large effect on the shape of the spectrum or the relative strengths of its features.
    During the chromatic phase, lensing-induced continuum shifts and spectral line distortions are visible in the ratios of the spectra. 
    Such chromatic distortions can affect the colors of the supernova.}
    \label{fig:lspec}
\end{figure*}

\begin{figure*}
	\centering
    \includegraphics[width=170mm]{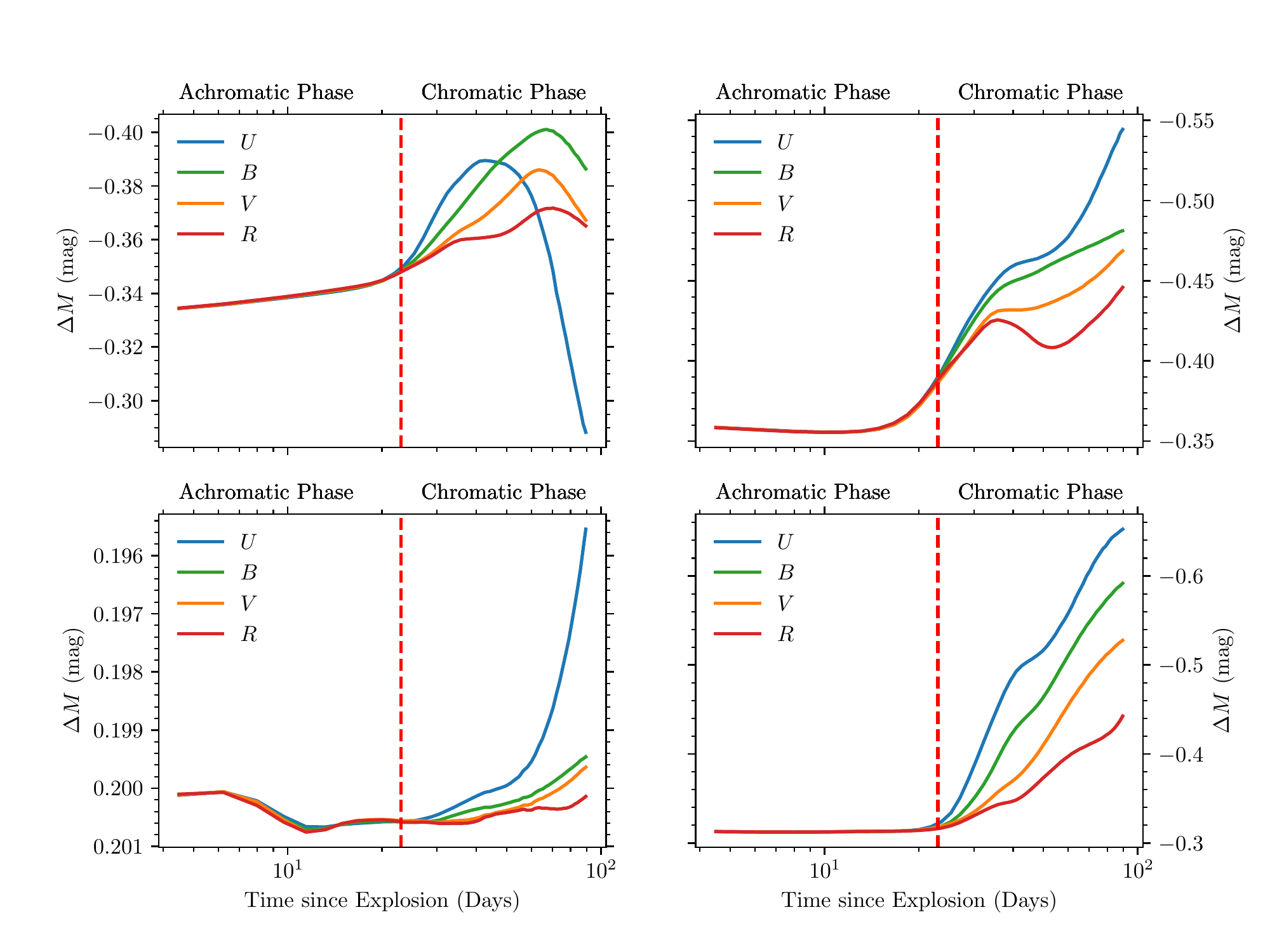}
    \caption{Four randomly chosen examples of broadband rest-frame difference light curves of model W7 computed by \MCcode.
    Each set of difference light curves has two distinct phases: an ``achromatic'' phase  in which $\Delta M$ is the same in all bands to within a few millimag, and a ``chromatic'' phase in which $\Delta M$ can vary significantly from band to band.
    }
   \label{fig:dlcs}
\end{figure*}

Example spectra and difference light curves of our microlensed \snia\ atmosphere appear in Figures \ref{fig:lspec} and \ref{fig:dlcs}, respectively; confidence regions of all $U-B$, $B-V$, $V-R$, and $R-I$ color curves produced by our simulation appear in Figure \ref{fig:ccs}.
The difference light curves give the microlensing amplification in magnitudes,
\begin{equation}
\Delta M(t) = -2.5 \log_{10} \left(\frac{L(t)}{\mu_0 U(t)}\right),
\end{equation}
where $U(t)$ and $L(t)$ are the unlensed  and observed fluxes of the supernova, respectively, and $\mu_0$ is the magnification in the absence of microlensing (i.e., if there were only macrolensing due to the lens galaxy). 
In the absence of microlensing, $\Delta M = 0$.

Each of these figures demonstrates that gl\snia\ microlensing has two phases, an ``achromatic'' phase, in which the microlensing magnification is the same at all wavelengths to a few millimag, followed by a  ``chromatic'' phase, in which the microlensing magnification varies strongly (and unpredictably) with wavelength.
The difference light curves show that the achromatic phase lasts roughly 3 rest-frame weeks after the explosion, transitioning to a chromatic phase between the time of peak brightness and the onset of the infrared secondary maximum.
During the achromatic phase, the light curves of gl\sneia\ can be deformed enough to bias time delay extraction; although $\Delta M$ is the same in all bands, it is not necessarily constant in time.

\subsection{Physics of Achromatic and Chromatic \snia\ Microlensing}
\begin{figure*}
	\centering
    \includegraphics[width=170mm]{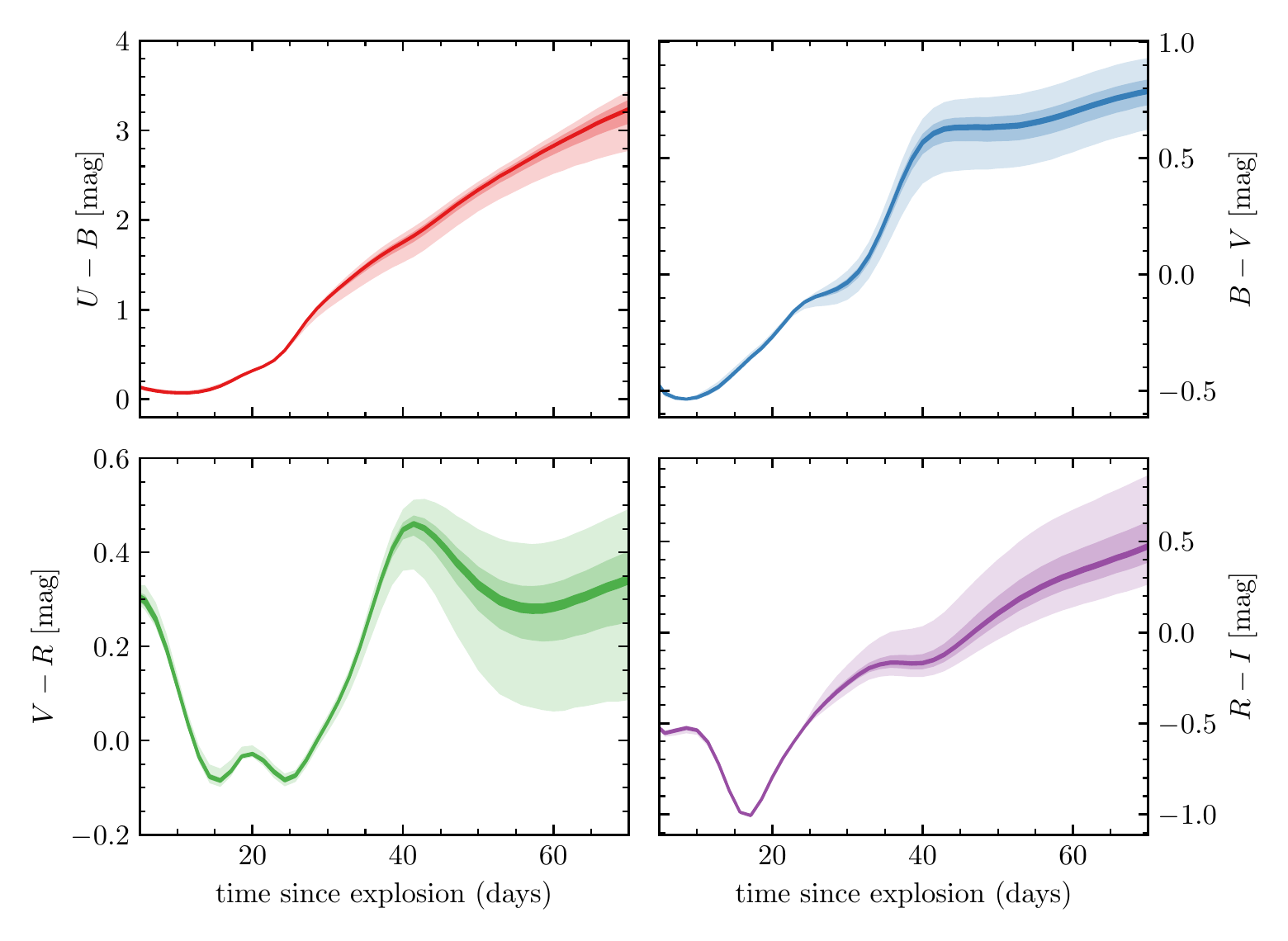}
    \caption{Rest-frame microlensed color curves of model W7.
    The intervals containing 68\%, 95\%, and 99\% of the 78,184 microlensed color curves described in Section \ref{sec:chrom} are plotted as progressively less opaque shaded regions.
    The color curves of the unlensed model are indistinguishable from the 68\% confidence regions of the microlensed models.   
    }
    \label{fig:ccs}
\end{figure*}

\begin{figure*}
	\centering
    \includegraphics[width=170mm]{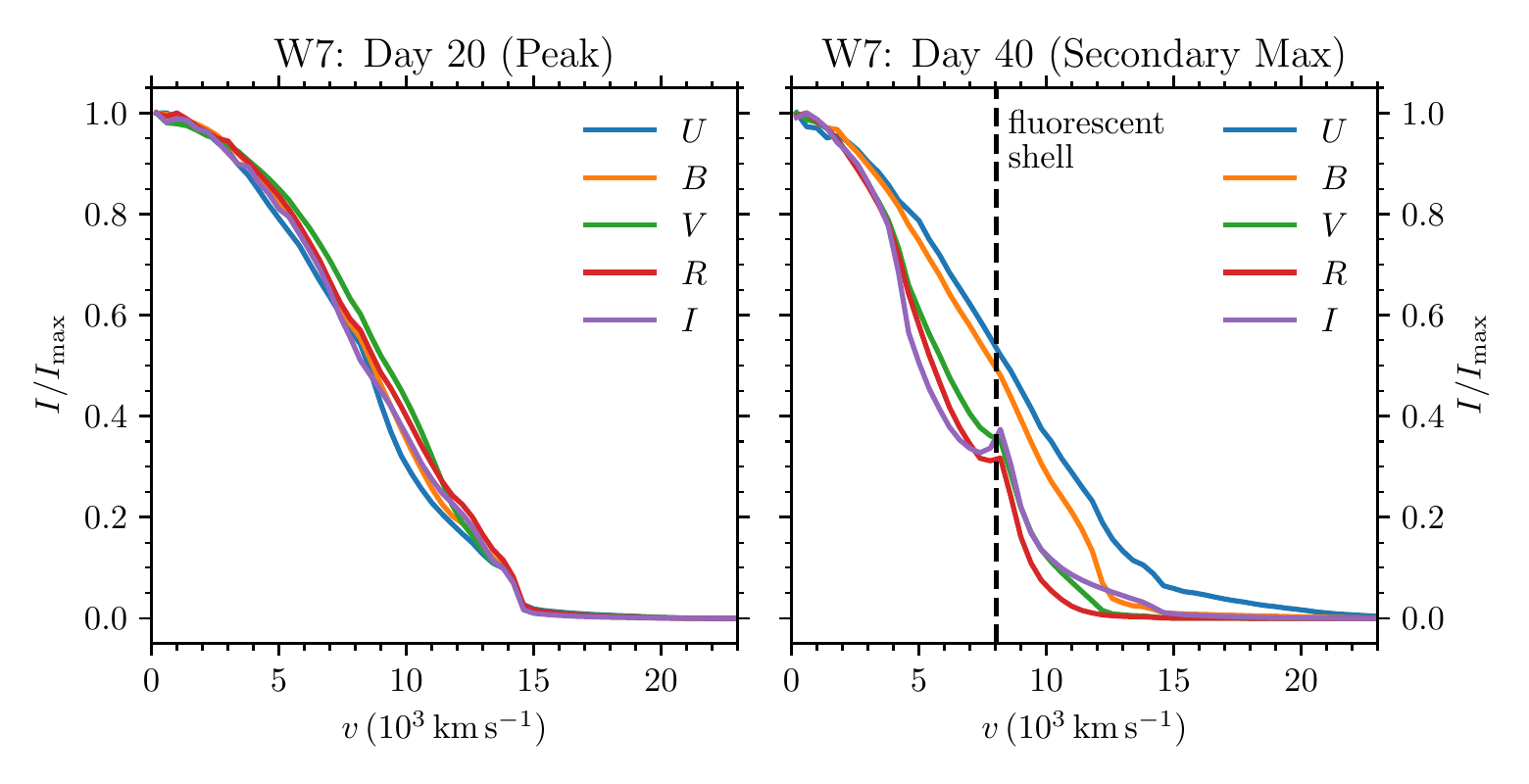}
    \caption{Normalized projected specific intensity profiles of model W7.
    Near peak, the specific intensity profiles in $UBVRI$ are similar, so microlensing is achromatic.
    At day 40, UV line blanketing and the ``fluorescent shell'' (in which Fe III $\rightarrow$ Fe II recombination occurs) causes different bands have different specific intensity profiles.
    As a result, microlensing is chromatic at this stage.}
    \label{fig:imap}
\end{figure*}

What is the physics responsible for the ``achromatic'' and ``chromatic'' phases of microlensing evident in Figures \ref{fig:dlcs} and \ref{fig:ccs}?
Figure \ref{fig:imap} shows the specific intensity profile $I_\lambda(v)$ of our unlensed model at 20 and 40 days after explosion, where $v$ is the velocity of the shell, equivalent to a radial variable (i.e., $P$).\footnote{As our model is spherically symmetric, $I_\lambda$ possesses no $\phi$-dependence.}
The left panel of the figure shows that near peak (20 days after explosion), the ratio $I_{X_1}(v)/I_{X_2}(v)$, where $X_1$ and $X_2$ are any two bands, is roughly  constant over all $v$. 
Thus the supernova near peak has a specific intensity profile that is independent of $v$ up to an overall normalization factor.
As a result, any magnification pattern $\mu(P, \phi)$ will not change its color.

However, after peak, the supernova expands and cools enough for some part of the atmosphere to reach a temperature of 7000K.
\cite{2bump} and \cite{kw07} note that this is the temperature at which Fe III recombines to Fe II, which presents a significantly higher opacity to blue and ultraviolet radiation than Fe III.
This line blanketing has the effect of enabling one to see redder emission from deeper in the supernova, while emission in the blue and the UV is pushed to larger radii.
Additionally, a ``fluorescent shell'' of iron recombination, which causes a peak in the redder bands in the specific intensity profile of the supernova, develops near the onset of the secondary maximum.
This shell is clearly visible in the righthand panel of Figure \ref{fig:imap}.
These two effects, line blanketing and a fluorescent shell, make the supernova's specific intensity ratio no longer spatially constant.
As a result, the supernova atmosphere becomes susceptible to chromatic fluctuations.

\section{The Effect of Microlensing on LSST Lensed Type Ia Supernova Yields}
\label{sec:salt}

Previous estimates of gl\snia\ yields, including those of \cite{om10}, \cite{quimby14}, and \cite{gn17}, modeled only the effects of macrolensing.
In this section, we present a novel method of identifying gl\sneia\ photometrically based on spectral template fitting.
We apply this technique to our simulated micro- and macrolensed gl\snia\ light curves to present a new estimate of gl\snia\ yields for LSST.

\subsection{Efficient Identification of Lensed Type Ia Supernovae with Spectral Template Fitting}
\label{sec:propmeth}

Our detection strategy rests on three observational facts.
First, normal \sneia\ are the brightest type of supernovae that have ever been observed to occur in elliptical galaxies \citep{maozreview}.
Second, the absolute magnitudes of normal \sneia\ in elliptical galaxies are remarkably homogenous, even without correcting for their colors or lightcurve shapes $(\sigma_M \about 0.4\ \mathrm{mag})$, with a component of the population being underluminous \citep{2011MNRAS.412.1441L}.
Finally, due to the sharp 4000\AA\ break in their spectra, elliptical galaxies tend to provide accurate photometric redshifts from large-scale multi-color galaxy surveys such as SDSS.

A high-cadence, wide-field imaging survey can leverage these facts to systematically search for strongly lensed \sneia\ in the following way.
First, by spatially cross-matching its list of supernova candidates with a catalog of elliptical galaxies for which secure photometric redshifts have been obtained,  supernovae that appear to be hosted by elliptical galaxies can be identified.
The hypothesis that one of these supernovae actually resides in its apparent host can be tested by fitting its broadband light curves with an \snia\  spectral template (as \sneia\ are the only types of supernovae that occur in ellipticals) fixed to the photometric redshift of the galaxy and constrained to obey $-18.5 > M_B > -20$, a liberal absolute magnitude range for \sneia, assuming a fiducial cosmology. 
If the transient is a lensed supernova at higher redshift, then the spectral template fit will fail catastrophically, as the supernova light curves will be strongly inconsistent with the redshift and brightness implied by the lens galaxy.

\subsection{Monte Carlo Simulation}
\label{sec:detectmc}

We use SALT2 \citep{salt2}, a parametrized \snia\ spectral template that is the \textit{de facto} standard tool to place \sneia\ on the Hubble diagram, to test this method.
The template possesses four parameters: $t_0$, $x_0$, $x_1$, and $c$, encoding a reference time, an overall SED normalization, a supernova ``stretch,'' and a color-law coefficient, respectively.
The flux of the template is given by 
\begin{equation}
F_\lambda(\lambda, t) = x_0 [M_0(\lambda, t) + x_1 M_1(\lambda, t)] \exp[c\,CL(\lambda)],
\end{equation}
where $M_0$ and $M_1$ are eigenspectra derived from a training sample of measured \snia\ spectra and $CL(\lambda)$ is the average color-correction law of the sample \citep[see][for details]{salt2}. 
The template aims to model the mean evolution of the SED sequence of \sneia\ and its variation with a few dominant components, including a time independent variation with color, whether it is intrinsic or due to extinction by dust in the host galaxy (or both).

We consider an LSST lensed supernova search in which the photometry is performed with a PSF that is artifically enlarged to blend the multiple images together into a single source.
We randomly assign each of the 37,100 simulated gl\sneia\ from Section \ref{sec:population} an LSST field from the nominal observing strategy (\texttt{minion\_1016}; LSST Science Collaborations in preparation).\footnote{\href{https://github.com/LSSTScienceCollaborations/ObservingStrategy}{https://github.com/LSSTScienceCollaborations/\\ObservingStrategy}}
We compute the phase- and wavelength- dependent magnification $\mu(\lambda, t)$ of each lensed image by placing its corresponding microlensed W7 SED into the rest frame, then  dividing each by the unlensed spectral sequence of the model.
We then generate the rest-frame spectral model for the image $F(\lambda, t)$ according to 
\begin{equation}
	\label{eq:hsiao}
	F(\lambda, t) = \mu(\lambda, t) H(\lambda, t),
\end{equation}
where $H(\lambda, t)$ is the \cite{hsiao} \snia\ spectral template.
We employ a warped Hsiao template rather than the microlensed W7 SEDs to mitigate uncertainties in the radiation transport.\footnote{The Hsiao template is an empirical, time-dependent SED model constructed from the observed spectra of many \sneia\, but it does not contain position dependent information (i.e., $P$, $\phi$), so to calculate $\mu(\lambda, t)$ theoretical models are needed.
The Hsiao template is more accurate than the theoretical model in the IR and after maximum light when NLTE effects become important.}
We then place the templates of each image at their correct redshifts, and we rescale and time-shift them to account for the macrolensing and time delays.
Finally, we draw a random time for the system (arbitrarily chosen to be the observer-frame time of rest-frame $B$-band maximum of the first image) from the 11 year period spanning 6 months before the beginning of the survey until 6 months following the end of the survey.

We realize broadband photometry of each blended gl\snia\ (summing the flux of each microlensed image) using the sky brightnesses, FWHMs, exposure times, observation times, and limiting magnitudes of the assigned field provided by \texttt{minion\_1016}, assuming the total area covered by the survey is 25,000 deg$^2$.
Starting from the first observation of the \snia, we fit the light curve with SALT2, fixed to the redshift of the lens galaxy (assumed to be known either as a photometric or spectroscopic redshift) and fixed to obey $-18.5 > M_B > -20$ at that redshift (effectively a constraint on $x_0$).
Additionally, we enforce bounds of $[-0.2, 0.2]$ on $c$ and $[-1, 1]$ on $x_1$, values characteristic of normal \sneia\ \citep{scalzo14a}.
We use the CERN minimization routine \texttt{MIGRAD}\ \citep{minuit} to fit the data.
If the light curve has at least one data point that is at least $5\sigma$ discrepant from the best fit and at least 4 data points with S/N $\geq 5$, then the object is marked ``detected."
If not, then the next observation is added and the process is repeated until the object is detected or all observations are added, resulting in a non-detection.

\begin{figure*}
	\centering
    \includegraphics[width=160mm]{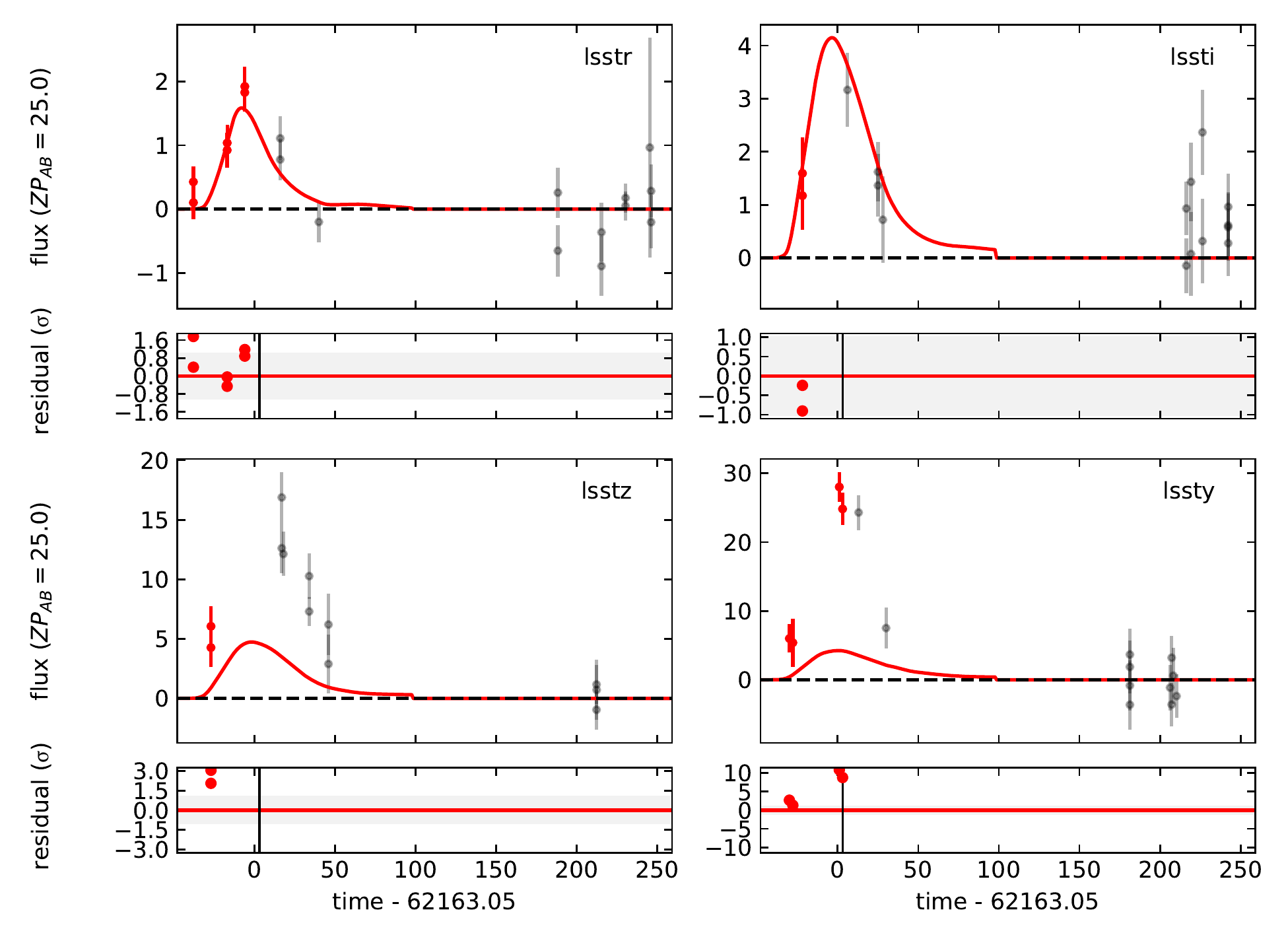}
    \caption{Detecting a $z_s=1.91, z_l=0.96$ LSST gl\snia\ with SALT2.
    The red data points show the ``current" light curve, and the red line shows the best fit SALT2 model fixed to the photometric redshift of the lens galaxy. 
    The gray points show future observations that are not included in this iteration of the fit.
    The black vertical line in the residual plots shows the date when the supernova is detected.
    Although the model fits the data well in the bluer bands, the high redshift of the source makes the data much brighter in the infrared than the model expects given the redshift of the lens.
Thus the object is detected shortly after peak due to $10\sigma$ discrepant points in $y$-band.}
    \label{fig:broadphot}
\end{figure*}

Figure \ref{fig:broadphot} shows an example of this procedure being used to detect one of our simulated gl\sneia\ at $z_s=1.91$.
The red data points show the ``current" light curve, and the red line shows the best fit model, fixed to $z_l=0.96$. 
Although the model fits the data well in the bluer bands, the high redshift of the source makes the data much brighter in the infrared than the model expects given the redshift of the lens.
Thus the object is detected shortly after peak due to $10\sigma$ discrepant points in $y$-band.

\subsection{Yields}
\label{sec:yields}
\begin{figure}
	\centering
    \includegraphics[width=1\columnwidth]{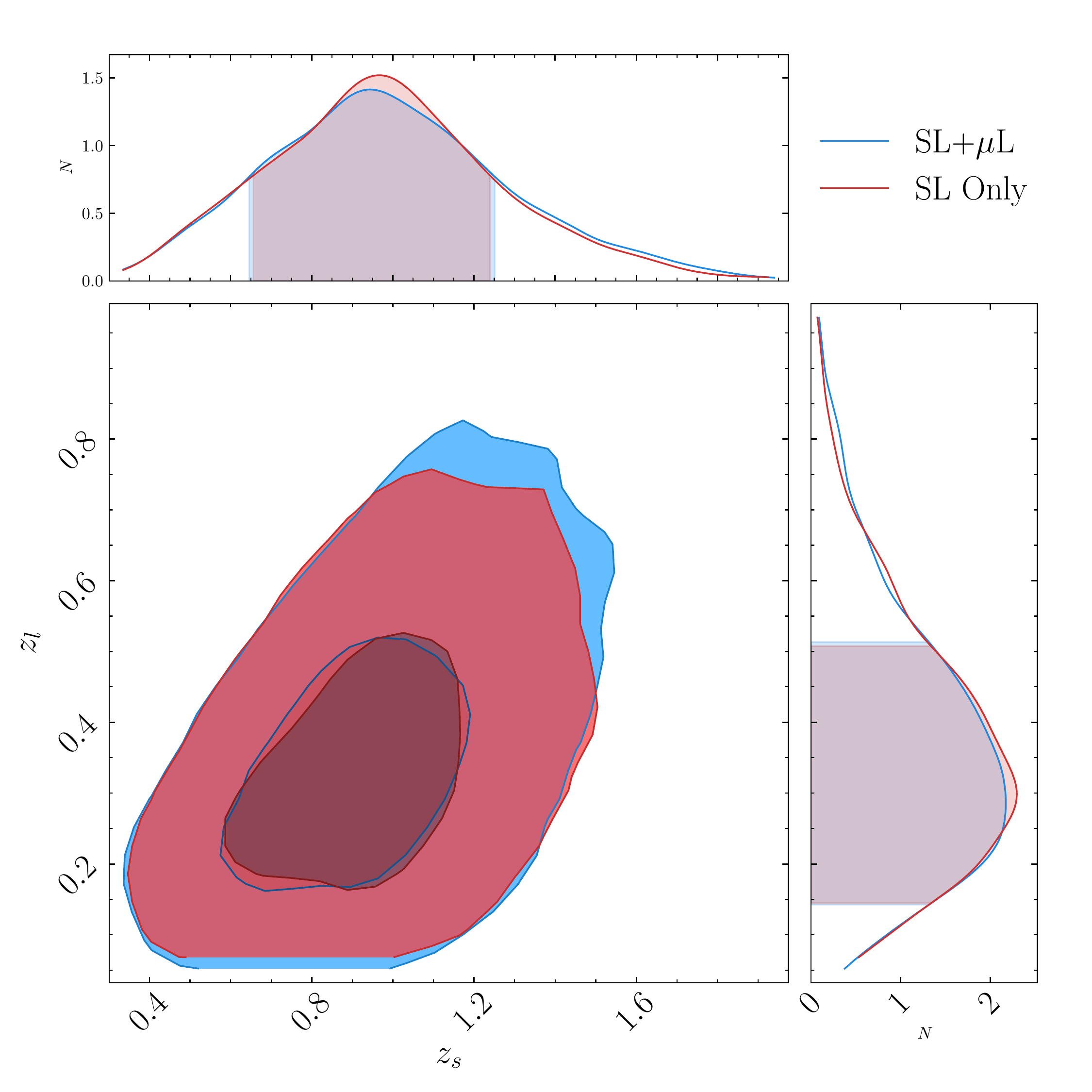}
    \caption{Source and lens redshift distributions of the gl\sneia\ with at least 4 data points with S/N $> 5$ detected with (blue) and without microlensing (red).
    The joint contours show 1 and 2$\sigma$.
    The marginal shaded regions show 1$\sigma$.}
    \label{fig:zdist}
\end{figure}

Our spectral template fitting approach to gl\snia\ identification delivers almost twice as many LSST gl\sneia\ than the method of \cite{gn17}.
In total, LSST should find $\sim$925 microlensed gl\sneia\ with the new method over the duration of its 10-year survey. 
This is almost identical to the case with no microlensing, which would yield 935 gl\sneia\ over the same period, with a nearly identical redshift distribution (see Figure \ref{fig:zdist}). 
This represents a major increase in the expected gl\snia\ yield for LSST, comparable to the number of expected lensed quasars \citep{om10}.

\begin{figure}
	\centering
	\includegraphics[width=1\columnwidth]{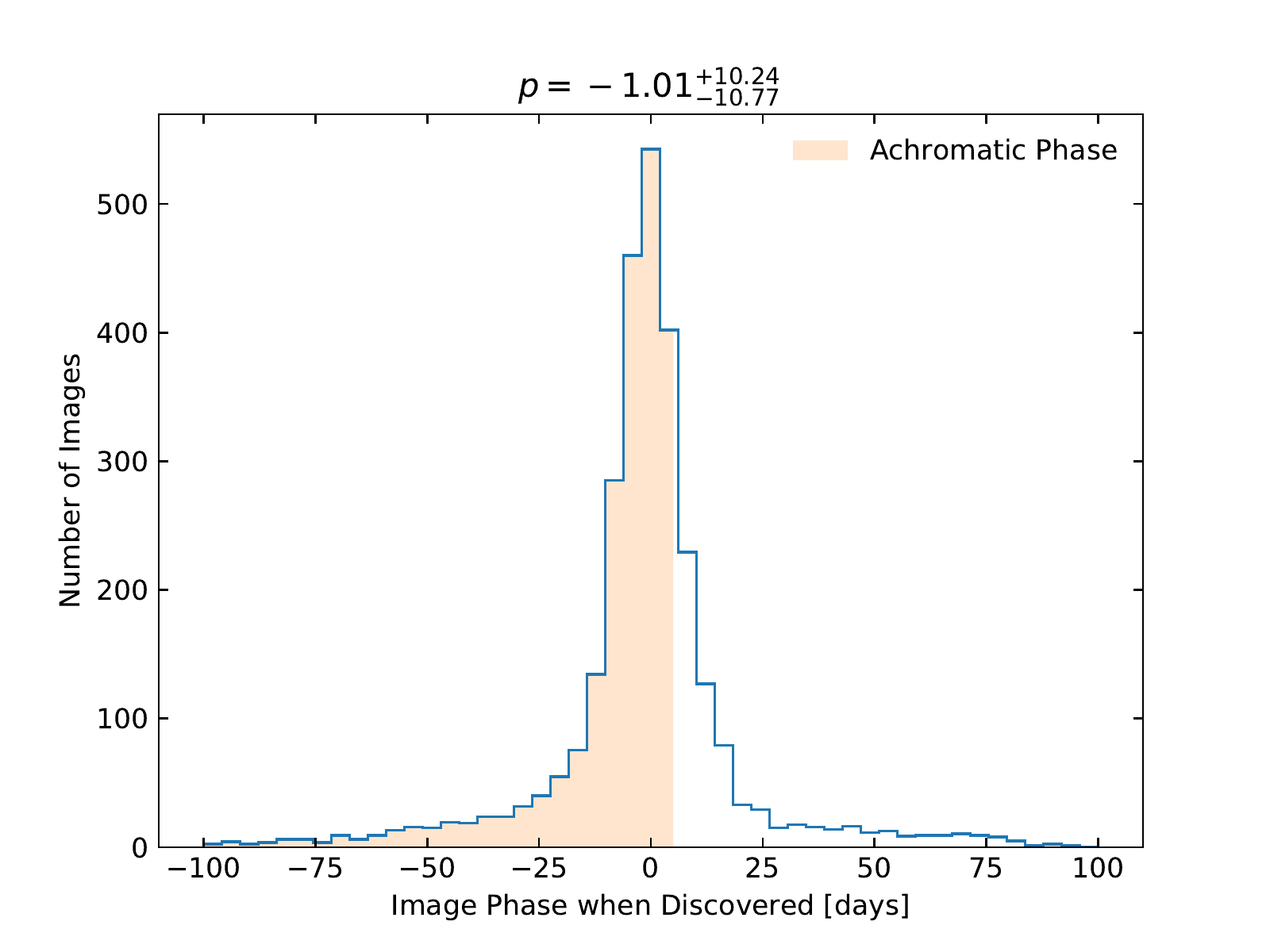}
    \caption{Phases of detected LSN images when they are discovered.
    Phases are given relative to each image, not to the peak of the total flux of the multiple blended images of SN.
    73\% of the images and 64\% of the image pairs are discovered during the achromatic phase.}
    \label{fig:phase}
\end{figure}

Figure \ref{fig:phase} shows the rest-frame phase distribution of discovered microlensed gl\snia\ images (a phase of 0 corresponds to peak brightness in $B$). 
The 68\% confidence interval of the image phase distribution is $-1.01^{+10.24}_{-10.77}$, so about half of the images should be discovered before peak brightness. 
73\% of the images and 64\% of the image pairs should be discovered during the achromatic phase.

\section{The Effect of Microlensing on Lensed Type Ia Supernova Time Delays}
\label{sec:dtmeth}
In Section \ref{sec:chrom}, we showed that microlensing introduces time- and wavelength-dependent fluctuations into the light curves of \sneia. 
In this section, we quantify the effect of these fluctuations on the time delays that can be extracted from simulated photometric observations of typical LSST gl\sneia, using as input the results of Section \ref{sec:yields}.
We demonstrate that microlensing can introduce time delay uncertainties of $\sim$4\% into the light curves of typical LSST gl\sneia, but that this number decreases to $\sim$1\% when achromatic-phase color curves of the same supernovae are used instead.

\subsection{Monte Carlo Simulations of Microlensing Time Delay Uncertainty}

\begin{figure*}
\centering
  \hspace{-4mm}\includegraphics[width=1\textwidth]{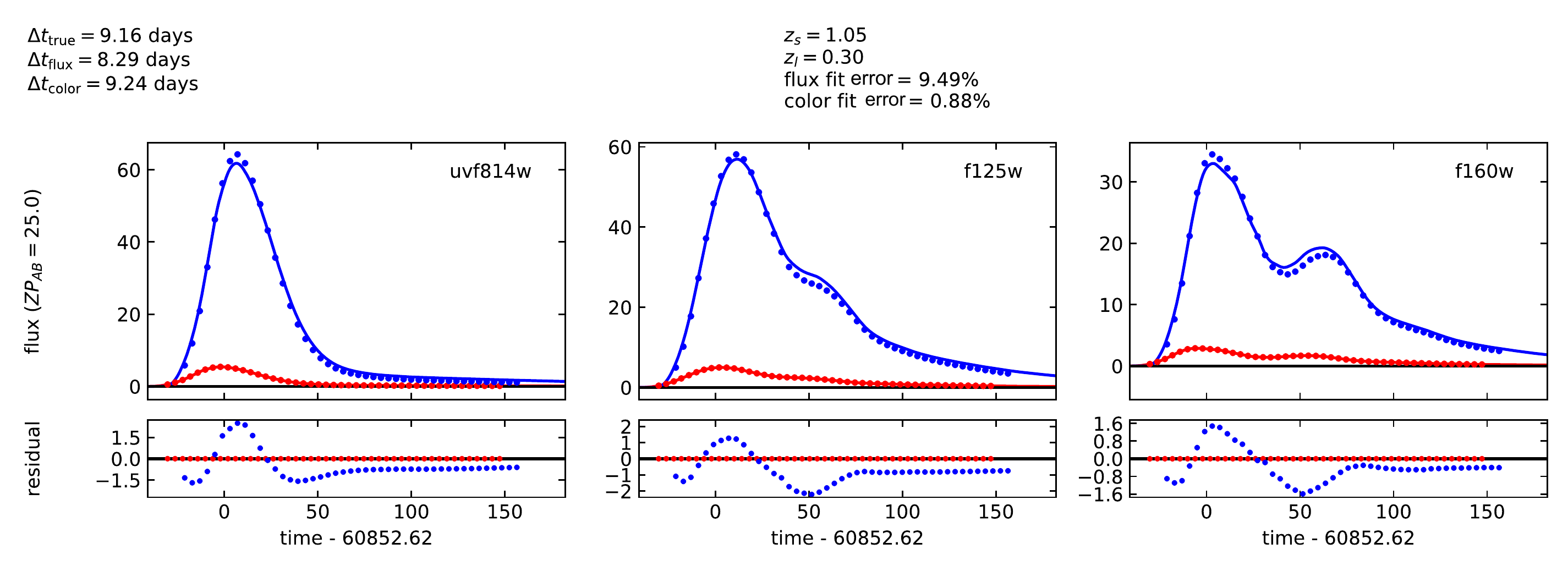}%
\par\medskip
  \includegraphics[width=1\textwidth]{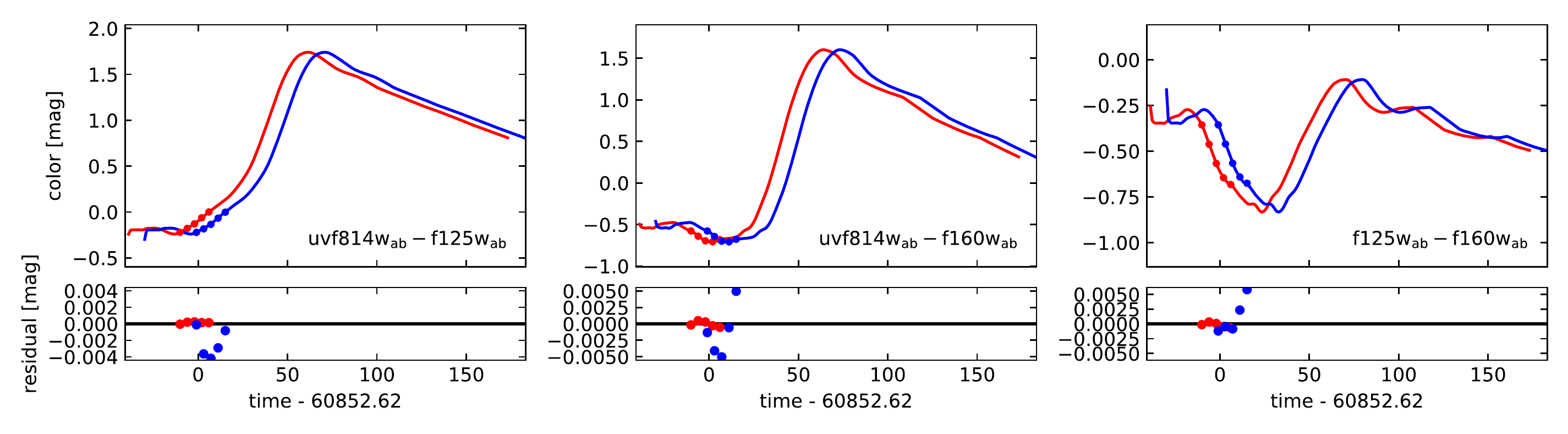}%
\caption{Fitting infinite S/N light curves and color curves (in the achromatic phase) of two images of a microlensed supernova with the unlensed Hsiao template to estimate time delay error.
Microlensing produces visible offsets in the features of the light curves, but the residuals show that during the achromatic phase (until a few weeks after peak brightness) the offsets are achromatic.
Thus when the color curves are fit in the achromatic phase, the uncertainty on the time delay is more than an order of magnitude smaller. 
}
\label{fig:fits}
\end{figure*}

\begin{figure}
	\centering
	\includegraphics[width=1\columnwidth]{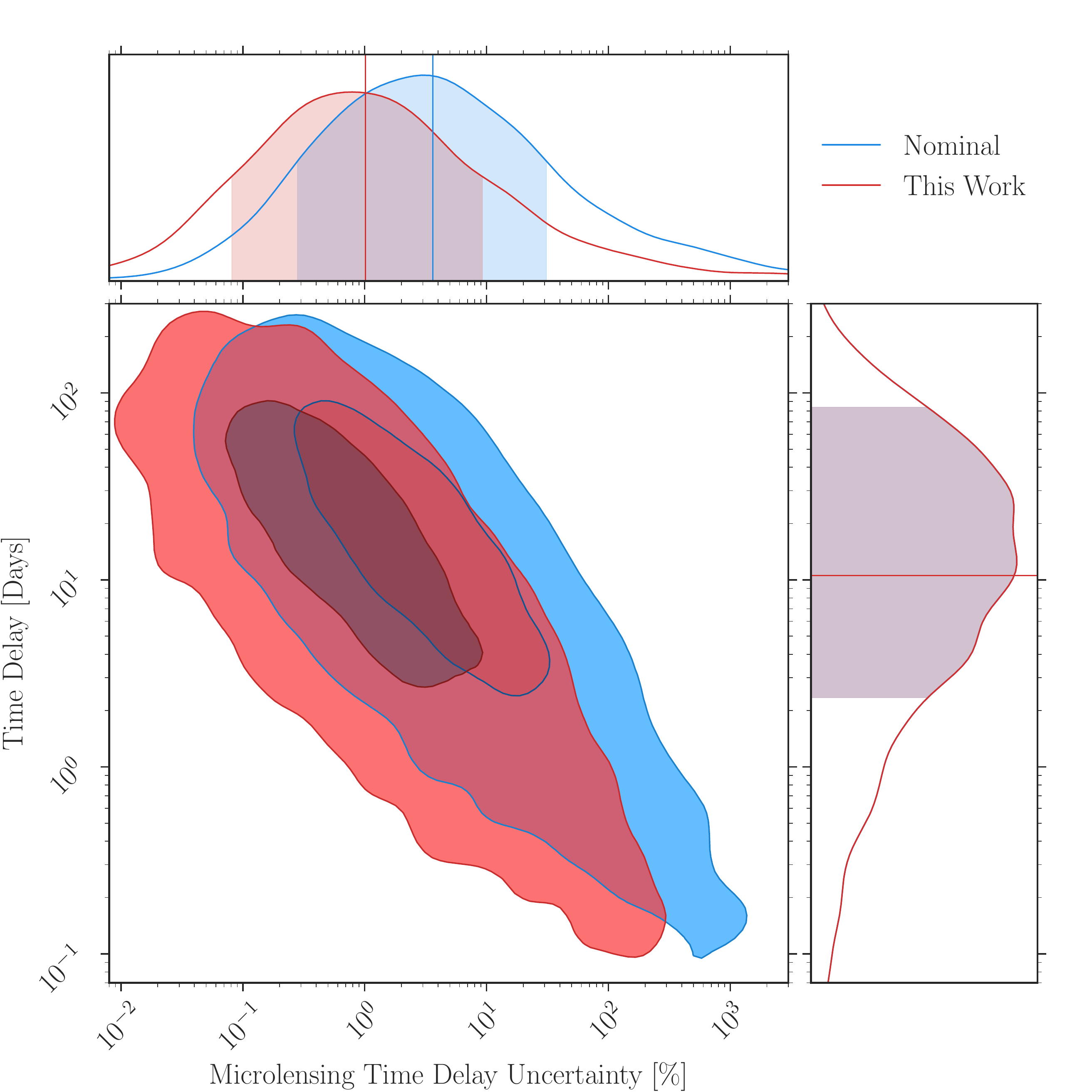}
    \caption{Joint distribution of the true time delay and the microlensing time delay uncertainty as a percentage of the true time delay for all discovered pairs of gl\snia\ images.
    The median time delay uncertainty is $4\%$ for light curves, but just 1$\%$ for color curve measurements in the achromatic phase.
    The median time delay that LSST will detect is 10 days.
    Contours show 1 and 2$\sigma$.
    Marginal shaded regions show 1$\sigma$.
    }
    \label{fig:prec}
\end{figure}

\begin{figure}
	\centering
    \includegraphics[width=1\columnwidth]{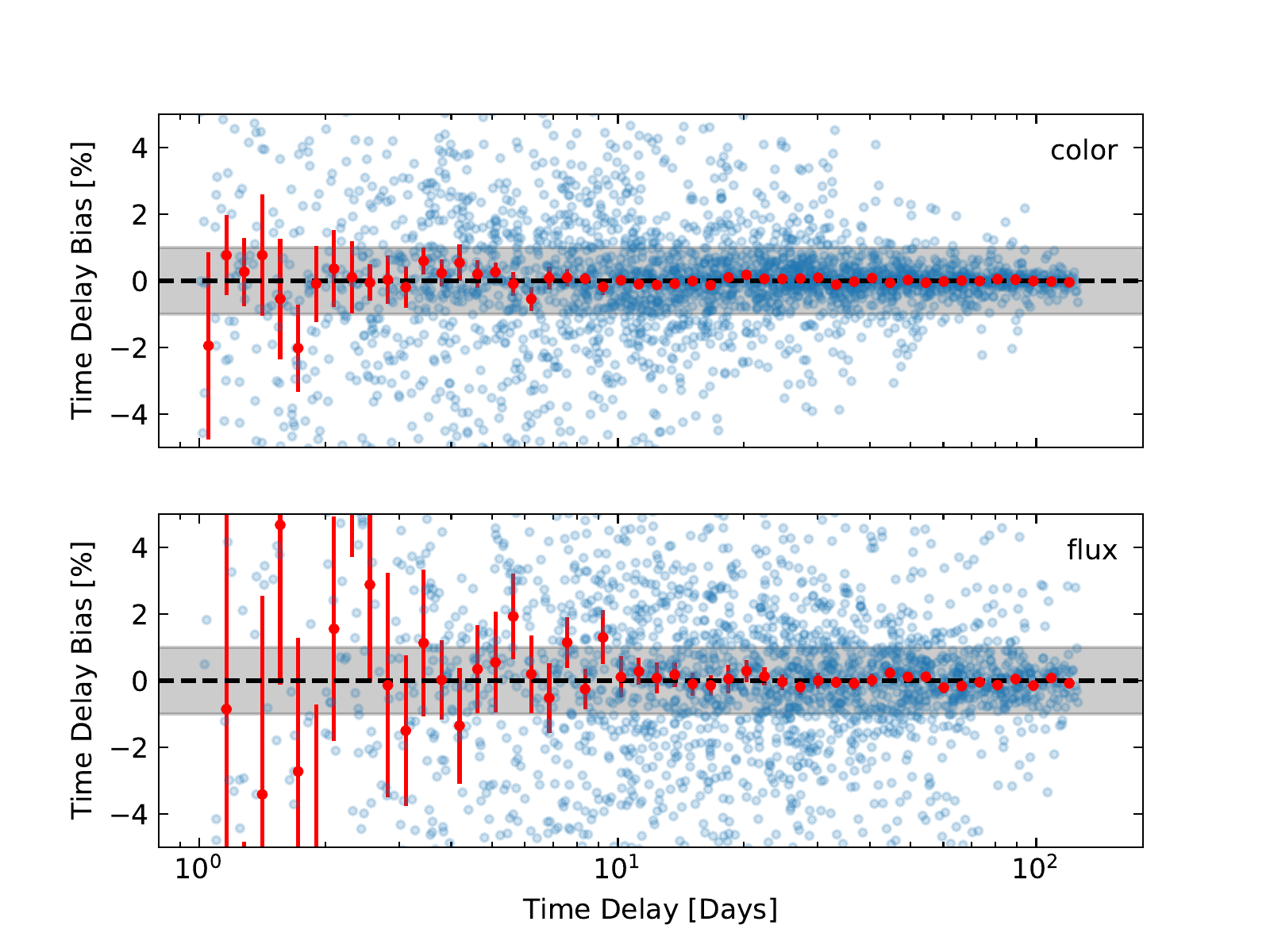}
    \caption{Bias as a function of true time delay for all discovered microlensed image pairs, fitting color curves in the achromatic phase (top panel) and entire light curves (bottom panel). 
    Blue points show individual fits; red points show bin averages. 
    }
    \label{fig:bias}
\end{figure}

Since many of the gl\sneia\ that will be discovered by LSST will require higher-spatial resolution follow-up observations to extract time delays, either with ground-based adaptive optics or space-based imaging, we simulate ``time-delay observations" with the Wide Field Camera 3 (WFC3) on the \textit{Hubble Space Telescope} (\hst).
The redshift distribution in Figure \ref{fig:zdist} implies that most LSST gl\sneia\ will be brightest in the IR, so we use the F814W, F125W, and F160W filters (roughly $I$, $J$, and $H$ bands, respectively) on WFC3  for this simulation. 

For each  of  the ``detected" microlensed gl\snia\ systems from Section \ref{sec:yields}, we realize 45 photometric observations of each image in F814W, F125W, and F160W with infinite signal to noise, with a uniform temporal spacing, spanning the light curve.
The spectral template for each image is the same as  the one used in Section \ref{sec:salt} (see Equation \ref{eq:hsiao}).
Using \texttt{MIGRAD}, we fit the realized light curves of each pair of images using the redshifted, unlensed spectral template $H(\lambda, t)$. 
For each fit, we estimate the fitted time delay as
\begin{equation}
\label{eq:td}
\dt = t_{0,2} - t_{0,1},
\end{equation}
where $t_{0,2}$ and $t_{0,1}$ are the fitted reference times of the second and first images, respectively.
We then measure the error on the fitted time delay as
\begin{equation}
	\label{eq:bias}
	\epsilon = \left|\frac{\Delta t - \Delta t'}{\Delta t'}\right|,
\end{equation}
where $\Delta t'$ is the true time delay of the pair of images.\footnote{N.B. If $\Delta t$ is distributed as a Gaussian with width $\sigma$, then the ensemble average of $\epsilon$ is only $0.79\sigma$.}

Next, we prune the photometric observations to the achromatic phase, masking all observations more than 5 rest-frame days from the date of $B$-band maximum. 
We then synthesize $\mathrm{F814W} - \mathrm{F125W}$, $\mathrm{F125W} - \mathrm{F160W}$, and $\mathrm{F814W} - \mathrm{F160W}$ color curves from the photometry.

We  apply the same fitting procedure to the color curves and estimate the fitted time delays and uncertainties using Equations \ref{eq:td} and \ref{eq:bias}.
Figure \ref{fig:fits} shows an example of the procedure being applied to the light curves and achromatic-phase color curves of two images of a supernova. 
While the light curve data show microlensing-induced offsets near peak brightness and the secondary maximum that produce a large time delay error of $\sim$10\%, the residuals show that these offsets are nearly the same in all bands during the achromatic phase, and thus the fits to the color curve have errors $<1$\%.

Figure \ref{fig:prec} shows the joint distribution of time delays and microlensing-induced time delay uncertainties for all detected pairs of images in our simulation.
We find that the median time delay of detected pairs of images is $\sim$10 days, and that the median microlensing-induced time delay uncertainty using light curve fits is 4\%, comparable to the current uncertainties on mass modeling.
However, this number drops down to 1\% ($\sim$2.5 hours on a 10 day time delay) when achromatic-phase color curves are used instead of light curves.

Figure \ref{fig:bias} shows the systematic microlensing bias on time delays from fitting light curves and color curves in the achromatic phase. 
The achromatic phase color curve fits are consistent with zero bias down to $\Delta t = 1$ day, while the light curve fits are consistent with zero bias down to $\Delta t = \sim$a few days. 
This result indicates that time delay bias from microlensing will not be a major systematic for cosmography with gl\sneia. 

\subsection{iPTF16geu}
iPTF16geu is the only gl\snia\ with resolved images that has been discovered to date. 
Here we consider its potential time delay precision and cosmological impact. 
Before it faded, multiwavelength follow-up observations of the event were obtained with a variable cadence using the Washington $C$, $g$, $r$, wide $I$, $Z$, $J$, and $H$ bands of WFC3 on \hst\ (DD 14862, PI: Goobar). 
The bluer bands were only observed sporadically, but the redder bands were observed with a roughly 4-day cadence.\footnote{A full description of the observations is available \href{ https://archive.stsci.edu/proposal\_search.php?mission=hst\&id=14862}{here}.}
Unfortunately, the observations were obtained well into the chromatic phase, and thus the light curves and color curves exhibit significant microlensing uncertainties. 
Final photometry of the event has not yet been produced, as not enough time has passed to take final reference images. 
The current best estimate of the time delay on this system is 35 hours \citep{goobar16}.
Based on Figure \ref{fig:prec}, we estimate the time delay uncertainty due to microlensing from this event to be $\sim$40\%. 
If the event were discovered earlier so that color curves during the achromatic phase could be constructed, then the microlensing time delay uncertainty would drop to $\sim$10\%.

\section{Conclusion}
\label{sec:conclusion}
In this article, we assessed the impact of microlensing on the yields and time delay precisions of LSST gl\sneia.
We presented microlensed broadband difference light curves and color curves of the well-understood \snia\ ejecta model W7 for 78,184 microlensing magnification patterns drawn from a realistic population model of gl\snia\ images.
We found that until shortly after peak brightness, the microlensing of \sneia\ is achromatic, and thus time delays from early-time color curves are less sensitive to microlensing than time delays from light curves.
We interpreted the achromaticity of microlensing before the onset of the secondary maximum as being due to UV line blanketing and the emergence of a fluorescent shell of Fe III $\rightarrow$ II recombination that alters the specific intensity profile of the ejecta as suggested by \cite{2bump} and \cite{kw07}.
We found that microlensing does not have a significant impact on gl\snia\ yields, but that they can be increased by a factor of $\sim$2 over the predictions of \cite{gn17} using a novel photometric detection techinque. 

Our \MCcode\ calculation of W7 represents the most detailed \snia\ spectrum synthesis calculation that has been used to investigate \snia\ microlensing to date, but it is inherently one-dimensional and makes a number of physical approximations for computational expediency.
It does not perfectly reproduce the observed colors of \sneia, especially at UV wavelengths, where line blanketing is strong and small differences in the underlying model can have pronounced effects.
Additionally, W7 is just one \snia\ model, and there is diversity in the \snia\ population.
However, we expect that this model captures the key physical behavior that leads to chromatic effects, and so we do not expect the results to change significantly with different one-dimensional models.
Although there is evidence from spectropolarimetry that global asymmetry in \sneia\ is very small \citep{specpol}, asymmetric explosion scenarios have not been ruled out by observations. 
In the future, it will be useful to assess whether asymmetric, multi-dimensional supernova models confirm the two phases of gl\snia\ microlensing identified in this work, or whether viewing angle effects become important.

Despite the complication of microlensing, time-delays can be robustly measured to sub-percent precision for many gl\sneia. 
By photometrically detecting the first image of a strongly lensed core-collapse supernova before light from the other images arrives, one can use the sharp shock-breakout light curve as a time delay indicator with precision $\sim$$(30\,\mathrm{min})(1+z_s) / \Delta t$.  
Such precision is difficult to achieve with lensed AGNs \citep{tewes13,bonvin17,tk17}, which also require a significantly longer observing campaign \citep{tdc}. 
This result provides a straightforward first step to inferring cosmological parameters with gl\sneia.
The latter steps of inferring the lens potential, and the effect of line-of-sight structures have already been implemented for lensed AGNs \citep[e.g.][]{suyu17,rusu17,wong17,bonvin17}, and the solutions from lensed AGNs should be directly portable to gl\sneia. 
Inferring the lens potential may even be easier than for AGNs as a more detailed reconstruction of the lensed \snia\ host should be possible once the supernova has faded. 
Those lensed \sneia\ with time-delays greater than a month are therefore golden lenses with which to measure $H_0$.
With a method for extracting precise time delays from these objects in hand, a renewed focus should be placed on the discovery and follow up of gl\sneia.

\appendix
\section{Radiation Transport Simulation}
\label{sec:sedona}
In this Appendix, we provide the details of the radiation transport simulations that we use to calculate the time-evolving \snia\ SED  in Section \ref{sec:sn}.
The \MCcode\ code \citep{sedona} is a time-dependent, multi-dimensional Monte Carlo radiative transfer code, designed to calculate the light curves, spectra and polarization of supernova explosion models.
Given a homologously expanding SN ejecta structure, \MCcode\ calculates the full time series of emergent spectra at high wavelength resolution.
In the present calculations, we employ a modified version of \MCcode\ that tags photons with their $P$ and $\phi$ values, thus we calculate $I_\lambda(P, \phi, t, \lambda)$. 
In addition to being fully time-dependent, the \MCcode\ calculation  accounts for the extendedness of continuum emitting regions in \snia\ atmospheres and the wavelength dependence of their location and extent.
It also explicitly accounts for light travel time across the atmosphere.
Broadband light curves are constructed by convolving the synthetic spectrum at each time with the appropriate instrumental throughputs.
\MCcode\ includes a detailed treatment of gamma-ray transfer to determine the instantaneous energy deposition rate from radioactive \Nifs\ and \Cofs\ decay.
Other decay chains that can change the composition of the ejecta, such as $\Crfe \rightarrow \Vfe \rightarrow \Tife$, are not treated.
Radiative heating and cooling rates are evaluated from Monte Carlo estimators, and the temperature structure of the ejecta is determined by iterating the model to thermal equilibrium.

Several significant approximations are made in our \MCcode\ simulation, notably the assumption of local thermodynamic equilibrium (LTE) in computing the atomic level populations.
In addition, bound-bound line transitions are treated using the expansion opacity formalism (implying the Sobolev approximation; \citealt{jeffery95}).
In this formalism, the opacities of spectral lines within a wavelength bin are represented in aggregate by a single effective opacity.
Although the \MCcode\ code is capable of a direct Monte Carlo treatment of NLTE line processes, due to computational constraints this functionality is not exploited here.
Instead, the line source functions are treated using an approximate two-level atom approach.
In the present calculations, we assume for simplicity that all lines are ``purely absorptive,'' i.e., in the two-level atom formalism the ratio of the probability of redistribution to pure scattering is taken to be $\epsilon_{\rm th} = 1$ for all lines.
In this case, the line source functions are given by the Planck function, consistent with our adoption of LTE level populations.

The numerical gridding in the present calculations was as follows:
\emph{spatial:} 100 equally spaced radial zones with a maximum velocity of $4 \times 10^4$~\kms;
\emph{temporal:} 459 time points beginning at day~1 and extending to day 100 with logarithmic spacing $\Delta \ log\,t = 0.175$;
\emph{wavelength:} covering the range 100-30,000~\AA\ with resolution of 10~\AA.
Extensive testing confirms the adequacy of this griding for the problem at hand.
Atomic line list data were taken from the Kurucz CD~23 line list \citep{Kurucz_Lines}, which contains nearly 500,000 lines.
$10^{10}$ photon packets were used for the calculation, which allowed for acceptable signal-to-noise in the synthetic broadband light curves, spectra, and velocity-dependent specific intensity profiles.

\section{Derivation of Equation 9}
\label{sec:geom}
\begin{figure}
	\centering
    \includegraphics[width=1\columnwidth]{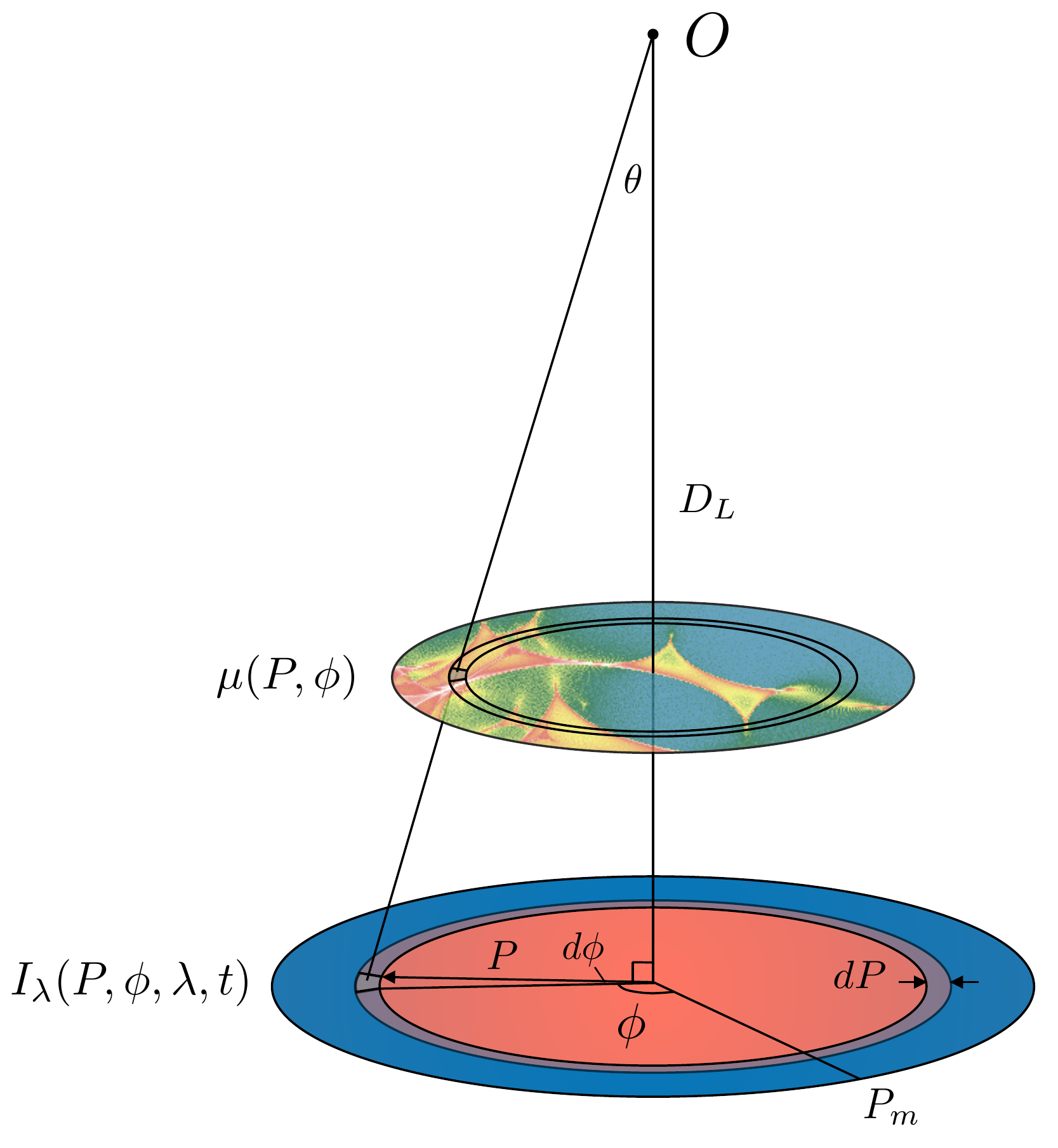}
    \caption{Integration geometry for Appendix \ref{sec:geom}.}
    \label{fig:geometry}
\end{figure}

The observed monochromatic flux density $F_\lambda$ of a source is obtained by setting up a small element of area $dA$ perpendicular to the line of sight at the location of the observer, and integrating the specific intensity of the field $I_\lambda$ in the direction normal to $dA$ over the solid angle subtended by the source \citep{rl},
\begin{equation}
\label{eq:def}
F_\lambda = \int I_\lambda \cos \theta d\Omega.
\end{equation}
In Equation \ref{eq:def}, $\theta$ is defined by $\tan \theta = P / D_L$, where $D_L$ is the luminosity distance from the observer to the closest point on the plane.\footnote{
The luminosity distance (not the angular diameter distance) is used here because the intrinsic luminosity of the source is known.}
From this definition we can construct the radial differential $dP$,
\begin{equation}
\label{eq:chvar}
dP = D_L \sec^2\theta\, d\theta.
\end{equation}
Using $d\Omega \equiv \sin\theta\,d\theta\,d\phi$, Equation \ref{eq:def} becomes
\begin{equation}
\label{eq:angular}
F_\lambda = \int^{2\pi}_0 \int^{\theta_m}_0 I_\lambda  \cos\theta \sin\theta \,d\theta\,d\phi,
\end{equation}
where $\theta_m$ is the maximum angular extent of the atmosphere.
Making the change of variables shown in Equation \ref{eq:chvar}, and using $\cos \theta = D_L / \sqrt{P^2 + D_L^2}$ and $\sin \theta = P / \sqrt{P^2 + D_L^2}$,  Equation \ref{eq:angular} becomes
\begin{equation}
\label{eq:lastref}
F_\lambda = \int^{2\pi}_0 \int^{P_m}_0 \frac{ PD_L^2 I_\lambda }{(P^2 + D_L^2)^2}\, dP\, d\phi.
\end{equation}
As we are in the $P \ll D_L$ limit, by Taylor expanding the denominator of the integrand of Equation \ref{eq:lastref} in powers of $(P/D_L)$ and keeping only first order terms, Equation \ref{eq:lastref} reduces to 
\begin{equation}
F_\lambda = D_L^{-2} \int^{2\pi}_0 \int_0^{P_m} I_\lambda \, P \, dP \, d\phi.
\label{eq:appx}
\end{equation}
Since lensing conserves surface brightness, the application of a spatially varying microlensing magnification pattern transforms $I_\lambda \rightarrow \mu(P, \phi) I_\lambda(P, \phi)$. 
Making this substitution in equation \ref{eq:appx}, we are left with Equation \ref{eq:lspec}.
A schematic of the integration geometry is presented in Figure \ref{fig:geometry}. 

\acknowledgements
DAG gratefully acknowledges Eric Linder for a careful reading of the manuscript and for thoughtful feedback.
Greg Aldering and Alex Kim also deserve thanks for feedback that improved the paper.
DAG, TEC, and PEN are members of the Large Synoptic Survey Telescope Dark Energy Science Collaboration (LSST-DESC), and they acknowledge Phil Marshall and the Strong Lensing working group for bringing them together so that this work could be launched.
DAG thanks Kyle Barbary for leading the development of \texttt{sncosmo} \citep{sncosmo}, which was essential to the analysis.
The authors acknowledge support from the DOE under grant DE-AC02-05CH11231, Analytical Modeling for Extreme-Scale Computing Environments. 
This research used resources of the National Energy Research Scientific Computing Center, a DOE Office of Science User Facility supported by the Office of Science of the U.S. Department of Energy under Contract No. DE-AC02-05CH11231 as well as the \texttt{savio} cluster at UC Berkeley.
Figures \ref{fig:mlsl}, \ref{fig:zdist}, and \ref{fig:prec} were rendered with code based on \texttt{ChainConsumer}\ \citep{chainconsumer}.
This research has made use of NASA's Astrophysics Data System. 
This research made use of Astropy, a community-developed core Python package for Astronomy \citep{astropy}.

\bibliography{ref}

\begin{thebibliography}{}
\expandafter\ifx\csname natexlab\endcsname\relax\def\natexlab#1{#1}\fi

\bibitem[{{Astropy Collaboration} {et~al.}(2013){Astropy Collaboration},
  {Robitaille}, {Tollerud}, {Greenfield}, {Droettboom}, {Bray}, {Aldcroft},
  {Davis}, {Ginsburg}, {Price-Whelan}, {Kerzendorf}, {Conley}, {Crighton},
  {Barbary}, {Muna}, {Ferguson}, {Grollier}, {Parikh}, {Nair}, {Unther},
  {Deil}, {Woillez}, {Conseil}, {Kramer}, {Turner}, {Singer}, {Fox}, {Weaver},
  {Zabalza}, {Edwards}, {Azalee Bostroem}, {Burke}, {Casey}, {Crawford},
  {Dencheva}, {Ely}, {Jenness}, {Labrie}, {Lim}, {Pierfederici}, {Pontzen},
  {Ptak}, {Refsdal}, {Servillat}, \& {Streicher}}]{astropy}
{Astropy Collaboration}, {Robitaille}, T.~P., {Tollerud}, E.~J., {et~al.} 2013,
  \aap, 558, A33

\bibitem[{{Bagherpour} {et~al.}(2006){Bagherpour}, {Branch}, \&
  {Kantowski}}]{bag06}
{Bagherpour}, H., {Branch}, D., \& {Kantowski}, R. 2006, \apj, 638, 946

\bibitem[{{Barbary} {et~al.}(2016){Barbary}, {Barclay}, {Biswas}, {Craig},
  {Feindt}, {Friesen}, {Goldstein}, {Jha}, {Rodney}, {Sofiatti}, {Thomas}, \&
  {Wood-Vasey}}]{sncosmo}
{Barbary}, K., {Barclay}, T., {Biswas}, R., {et~al.} 2016, {SNCosmo: Python
  library for supernova cosmology}, Astrophysics Source Code Library,
  ascl:1611.017

\bibitem[{{Bezanson} {et~al.}(2011){Bezanson}, {van Dokkum}, {Franx},
  {Brammer}, {Brinchmann}, {Kriek}, {Labb{\'e}}, {Quadri}, {Rix}, {van de
  Sande}, {Whitaker}, \& {Williams}}]{bezanson11}
{Bezanson}, R., {van Dokkum}, P.~G., {Franx}, M., {et~al.} 2011, \apjl, 737,
  L31

\bibitem[{{Bonvin} {et~al.}(2016){Bonvin}, {Tewes}, {Courbin}, {Kuntzer},
  {Sluse}, \& {Meylan}}]{bonvin16}
{Bonvin}, V., {Tewes}, M., {Courbin}, F., {et~al.} 2016, \aap, 585, A88

\bibitem[{{Bonvin} {et~al.}(2017){Bonvin}, {Courbin}, {Suyu}, {Marshall},
  {Rusu}, {Sluse}, {Tewes}, {Wong}, {Collett}, {Fassnacht}, {Treu}, {Auger},
  {Hilbert}, {Koopmans}, {Meylan}, {Rumbaugh}, {Sonnenfeld}, \&
  {Spiniello}}]{bonvin17}
{Bonvin}, V., {Courbin}, F., {Suyu}, S.~H., {et~al.} 2017, \mnras, 465, 4914

\bibitem[{{Chae}(2003)}]{chae03}
{Chae}, K.-H. 2003, \mnras, 346, 746

\bibitem[{{Chae}(2007)}]{chae07}
---. 2007, \apjl, 658, L71

\bibitem[{{Chang} \& {Refsdal}(1979)}]{ml}
{Chang}, K., \& {Refsdal}, S. 1979, \nat, 282, 561

\bibitem[{{Choi} {et~al.}(2007){Choi}, {Park}, \& {Vogeley}}]{choi07}
{Choi}, Y.-Y., {Park}, C., \& {Vogeley}, M.~S. 2007, \apj, 658, 884

\bibitem[{{Collett} \& {Auger}(2014)}]{collett14}
{Collett}, T.~E., \& {Auger}, M.~W. 2014, \mnras, 443, 969

\bibitem[{{Collett} \& {Cunnington}(2016)}]{collettcunnington16}
{Collett}, T.~E., \& {Cunnington}, S.~D. 2016, \mnras, 462, 3255

\bibitem[{{Collett} {et~al.}(2013){Collett}, {Marshall}, {Auger}, {Hilbert},
  {Suyu}, {Greene}, {Treu}, {Fassnacht}, {Koopmans}, {Brada{\v c}}, \&
  {Blandford}}]{collett13}
{Collett}, T.~E., {Marshall}, P.~J., {Auger}, M.~W., {et~al.} 2013, \mnras,
  432, 679

\bibitem[{{Dobler} \& {Keeton}(2006)}]{dk06}
{Dobler}, G., \& {Keeton}, C.~R. 2006, \apj, 653, 1391

\bibitem[{{Fassnacht} {et~al.}(2011){Fassnacht}, {Koopmans}, \&
  {Wong}}]{fassnacht11}
{Fassnacht}, C.~D., {Koopmans}, L.~V.~E., \& {Wong}, K.~C. 2011, \mnras, 410,
  2167

\bibitem[{{Goldstein} \& {Nugent}(2017)}]{gn17}
{Goldstein}, D.~A., \& {Nugent}, P.~E. 2017, \apjl, 834, L5

\bibitem[{{Goobar} {et~al.}(2017){Goobar}, {Amanullah}, {Kulkarni}, {Nugent},
  {Johansson}, {Steidel}, {Law}, {M{\"o}rtsell}, {Quimby}, {Blagorodnova},
  {Brandeker}, {Cao}, {Cooray}, {Ferretti}, {Fremling}, {Hangard}, {Kasliwal},
  {Kupfer}, {Lunnan}, {Masci}, {Miller}, {Nayyeri}, {Neill}, {Ofek},
  {Papadogiannakis}, {Petrushevska}, {Ravi}, {Sollerman}, {Sullivan}, {Taddia},
  {Walters}, {Wilson}, {Yan}, \& {Yaron}}]{goobar16}
{Goobar}, A., {Amanullah}, R., {Kulkarni}, S.~R., {et~al.} 2017, Science, 356,
  291

\bibitem[{{Guy} {et~al.}(2007){Guy}, {Astier}, {Baumont}, {Hardin}, {Pain},
  {Regnault}, {Basa}, {Carlberg}, {Conley}, {Fabbro}, {Fouchez}, {Hook},
  {Howell}, {Perrett}, {Pritchet}, {Rich}, {Sullivan}, {Antilogus}, {Aubourg},
  {Bazin}, {Bronder}, {Filiol}, {Palanque-Delabrouille}, {Ripoche}, \&
  {Ruhlmann-Kleider}}]{salt2}
{Guy}, J., {Astier}, P., {Baumont}, S., {et~al.} 2007, \aap, 466, 11

\bibitem[{Hinton(2016)}]{chainconsumer}
Hinton, S. 2016, {JOSS}, 1, doi:10.21105/joss.00045

\bibitem[{{Holder} \& {Schechter}(2003)}]{holder03}
{Holder}, G.~P., \& {Schechter}, P.~L. 2003, \apj, 589, 688

\bibitem[{{Hsiao} {et~al.}(2007){Hsiao}, {Conley}, {Howell}, {Sullivan},
  {Pritchet}, {Carlberg}, {Nugent}, \& {Phillips}}]{hsiao}
{Hsiao}, E.~Y., {Conley}, A., {Howell}, D.~A., {et~al.} 2007, \apj, 663, 1187

\bibitem[{{James} \& {Roos}(1975)}]{minuit}
{James}, F., \& {Roos}, M. 1975, Computer Physics Communications, 10, 343

\bibitem[{{Jeffery}(1995)}]{jeffery95}
{Jeffery}, D.~J. 1995, \aap, 299, 770

\bibitem[{{Kasen}(2006)}]{2bump}
{Kasen}, D. 2006, \apj, 649, 939

\bibitem[{{Kasen} {et~al.}(2006){Kasen}, {Thomas}, \& {Nugent}}]{sedona}
{Kasen}, D., {Thomas}, R.~C., \& {Nugent}, P. 2006, \apj, 651, 366

\bibitem[{{Kasen} \& {Woosley}(2007)}]{kw07}
{Kasen}, D., \& {Woosley}, S.~E. 2007, \apj, 656, 661

\bibitem[{{Kayser} {et~al.}(1986){Kayser}, {Refsdal}, \& {Stabell}}]{cml}
{Kayser}, R., {Refsdal}, S., \& {Stabell}, R. 1986, \aap, 166, 36

\bibitem[{{Keeton} {et~al.}(1997){Keeton}, {Kochanek}, \& {Seljak}}]{keeton97}
{Keeton}, C.~R., {Kochanek}, C.~S., \& {Seljak}, U. 1997, \apj, 482, 604

\bibitem[{{Kochanek}(1991)}]{kochanek91}
{Kochanek}, C.~S. 1991, \apj, 373, 354

\bibitem[{{Koopmans} {et~al.}(2009){Koopmans}, {Bolton}, {Treu}, {Czoske},
  {Auger}, {Barnab{\`e}}, {Vegetti}, {Gavazzi}, {Moustakas}, \&
  {Burles}}]{koopmans09}
{Koopmans}, L.~V.~E., {Bolton}, A., {Treu}, T., {et~al.} 2009, \apjl, 703, L51

\bibitem[{{Kormann} {et~al.}(1994){Kormann}, {Schneider}, \&
  {Bartelmann}}]{kormann94}
{Kormann}, R., {Schneider}, P., \& {Bartelmann}, M. 1994, \aap, 284, 285

\bibitem[{{Kurucz} \& {Bell}(1995)}]{Kurucz_Lines}
{Kurucz}, R.~L., \& {Bell}, B. 1995, {Atomic line list}

\bibitem[{{Li} {et~al.}(2011){Li}, {Leaman}, {Chornock}, {Filippenko},
  {Poznanski}, {Ganeshalingam}, {Wang}, {Modjaz}, {Jha}, {Foley}, \&
  {Smith}}]{2011MNRAS.412.1441L}
{Li}, W., {Leaman}, J., {Chornock}, R., {et~al.} 2011, \mnras, 412, 1441

\bibitem[{{Liao} {et~al.}(2015){Liao}, {Treu}, {Marshall}, {Fassnacht},
  {Rumbaugh}, {Dobler}, {Aghamousa}, {Bonvin}, {Courbin}, {Hojjati}, {Jackson},
  {Kashyap}, {Rathna Kumar}, {Linder}, {Mandel}, {Meng}, {Meylan}, {Moustakas},
  {Prabhu}, {Romero-Wolf}, {Shafieloo}, {Siemiginowska}, {Stalin}, {Tak},
  {Tewes}, \& {van Dyk}}]{tdc}
{Liao}, K., {Treu}, T., {Marshall}, P., {et~al.} 2015, \apj, 800, 11

\bibitem[{{Linder}(2004)}]{linder04}
{Linder}, E.~V. 2004, \prd, 70, 043534

\bibitem[{{Linder}(2011)}]{linder11}
---. 2011, \prd, 84, 123529

\bibitem[{{LSST Science Collaboration} {et~al.}(2009){LSST Science
  Collaboration}, {Abell}, {Allison}, {Anderson}, {Andrew}, {Angel}, {Armus},
  {Arnett}, {Asztalos}, {Axelrod}, \& et~al.}]{lsst}
{LSST Science Collaboration}, {Abell}, P.~A., {Allison}, J., {et~al.} 2009,
  ArXiv e-prints, arXiv:0912.0201

\bibitem[{{Maoz} {et~al.}(2014){Maoz}, {Mannucci}, \& {Nelemans}}]{maozreview}
{Maoz}, D., {Mannucci}, F., \& {Nelemans}, G. 2014, \araa, 52, 107

\bibitem[{{McCully} {et~al.}(2017){McCully}, {Keeton}, {Wong}, \&
  {Zabludoff}}]{mccully17}
{McCully}, C., {Keeton}, C.~R., {Wong}, K.~C., \& {Zabludoff}, A.~I. 2017,
  \apj, 836, 141

\bibitem[{Moore \& Hewitt(1996)}]{moore96}
Moore, C.~B., \& Hewitt, J.~N. 1996, Prospects for the Detection of
  Microlensing Time Delays, ed. C.~S. Kochanek \& J.~N. Hewitt (Dordrecht:
  Springer Netherlands), 279--280

\bibitem[{{More} {et~al.}(2017){More}, {Suyu}, {Oguri}, {More}, \&
  {Lee}}]{more16}
{More}, A., {Suyu}, S.~H., {Oguri}, M., {More}, S., \& {Lee}, C.-H. 2017,
  \apjl, 835, L25

\bibitem[{{Nomoto} {et~al.}(1984){Nomoto}, {Thielemann}, \& {Yokoi}}]{w7}
{Nomoto}, K., {Thielemann}, F.-K., \& {Yokoi}, K. 1984, \apj, 286, 644

\bibitem[{{Oguri}(2010)}]{glafic}
{Oguri}, M. 2010, \pasj, 62, 1017

\bibitem[{{Oguri} \& {Kawano}(2003)}]{oguri03}
{Oguri}, M., \& {Kawano}, Y. 2003, \mnras, 338, L25

\bibitem[{{Oguri} \& {Marshall}(2010)}]{om10}
{Oguri}, M., \& {Marshall}, P.~J. 2010, \mnras, 405, 2579

\bibitem[{{Oguri} {et~al.}(2008){Oguri}, {Inada}, {Strauss}, {Kochanek},
  {Richards}, {Schneider}, {Becker}, {Fukugita}, {Gregg}, {Hall}, {Hennawi},
  {Johnston}, {Kayo}, {Keeton}, {Pindor}, {Shin}, {Turner}, {White}, {York},
  {Anderson}, {Bahcall}, {Brunner}, {Burles}, {Castander}, {Chiu},
  {Clocchiatti}, {Eisenstein}, {Frieman}, {Kawano}, {Lupton}, {Morokuma},
  {Rix}, {Scranton}, \& {Sheldon}}]{oguri08}
{Oguri}, M., {Inada}, N., {Strauss}, M.~A., {et~al.} 2008, \aj, 135, 512

\bibitem[{{Oguri} {et~al.}(2012){Oguri}, {Inada}, {Strauss}, {Kochanek},
  {Kayo}, {Shin}, {Morokuma}, {Richards}, {Rusu}, {Frieman}, {Fukugita},
  {Schneider}, {York}, {Bahcall}, \& {White}}]{oguri12}
---. 2012, \aj, 143, 120

\bibitem[{{Perlmutter} {et~al.}(1999){Perlmutter}, {Aldering}, {Goldhaber},
  {Knop}, {Nugent}, {Castro}, {Deustua}, {Fabbro}, {Goobar}, {Groom}, {Hook},
  {Kim}, {Kim}, {Lee}, {Nunes}, {Pain}, {Pennypacker}, {Quimby}, {Lidman},
  {Ellis}, {Irwin}, {McMahon}, {Ruiz-Lapuente}, {Walton}, {Schaefer}, {Boyle},
  {Filippenko}, {Matheson}, {Fruchter}, {Panagia}, {Newberg}, {Couch}, \&
  {Project}}]{saul}
{Perlmutter}, S., {Aldering}, G., {Goldhaber}, G., {et~al.} 1999, \apj, 517,
  565

\bibitem[{{Planck Collaboration} {et~al.}(2016){Planck Collaboration}, {Ade},
  {Aghanim}, {Arnaud}, {Ashdown}, {Aumont}, {Baccigalupi}, {Banday},
  {Barreiro}, {Bartlett}, \& et~al.}]{planck15}
{Planck Collaboration}, {Ade}, P.~A.~R., {Aghanim}, N., {et~al.} 2016, \aap,
  594, A13

\bibitem[{{Quimby} {et~al.}(2014){Quimby}, {Oguri}, {More}, {More}, {Moriya},
  {Werner}, {Tanaka}, {Folatelli}, {Bersten}, {Maeda}, \& {Nomoto}}]{quimby14}
{Quimby}, R.~M., {Oguri}, M., {More}, A., {et~al.} 2014, Science, 344, 396

\bibitem[{{Riess} {et~al.}(1998){Riess}, {Filippenko}, {Challis},
  {Clocchiatti}, {Diercks}, {Garnavich}, {Gilliland}, {Hogan}, {Jha},
  {Kirshner}, {Leibundgut}, {Phillips}, {Reiss}, {Schmidt}, {Schommer},
  {Smith}, {Spyromilio}, {Stubbs}, {Suntzeff}, \& {Tonry}}]{adam}
{Riess}, A.~G., {Filippenko}, A.~V., {Challis}, P., {et~al.} 1998, \aj, 116,
  1009

\bibitem[{{Riess} {et~al.}(2016){Riess}, {Macri}, {Hoffmann}, {Scolnic},
  {Casertano}, {Filippenko}, {Tucker}, {Reid}, {Jones}, {Silverman},
  {Chornock}, {Challis}, {Yuan}, {Brown}, \& {Foley}}]{riess16}
{Riess}, A.~G., {Macri}, L.~M., {Hoffmann}, S.~L., {et~al.} 2016, \apj, 826, 56

\bibitem[{{Rusu} {et~al.}(2017){Rusu}, {Fassnacht}, {Sluse}, {Hilbert}, {Wong},
  {Huang}, {Suyu}, {Collett}, {Marshall}, {Treu}, \& {Koopmans}}]{rusu17}
{Rusu}, C.~E., {Fassnacht}, C.~D., {Sluse}, D., {et~al.} 2017, \mnras, 467,
  4220

\bibitem[{{Rybicki} \& {Lightman}(1979)}]{rl}
{Rybicki}, G.~B., \& {Lightman}, A.~P. 1979, {Radiative processes in
  astrophysics}

\bibitem[{{Salpeter}(1955)}]{salpeter}
{Salpeter}, E.~E. 1955, \apj, 121, 161

\bibitem[{{Scalzo} {et~al.}(2014){Scalzo}, {Aldering}, {Antilogus}, {Aragon},
  {Bailey}, {Baltay}, {Bongard}, {Buton}, {Cellier-Holzem}, {Childress},
  {Chotard}, {Copin}, {Fakhouri}, {Gangler}, {Guy}, {Kim}, {Kowalski},
  {Kromer}, {Nordin}, {Nugent}, {Paech}, {Pain}, {Pecontal}, {Pereira},
  {Perlmutter}, {Rabinowitz}, {Rigault}, {Runge}, {Saunders}, {Sim}, {Smadja},
  {Tao}, {Taubenberger}, {Thomas}, {Weaver}, \& {Nearby Supernova
  Factory}}]{scalzo14a}
{Scalzo}, R., {Aldering}, G., {Antilogus}, P., {et~al.} 2014, \mnras, 440, 1498

\bibitem[{{Schneider} \& {Weiss}(1987)}]{schneiderweiss87}
{Schneider}, P., \& {Weiss}, A. 1987, \aap, 171, 49

\bibitem[{{Sheth} {et~al.}(2003){Sheth}, {Bernardi}, {Schechter}, {Burles},
  {Eisenstein}, {Finkbeiner}, {Frieman}, {Lupton}, {Schlegel}, {Subbarao},
  {Shimasaku}, {Bahcall}, {Brinkmann}, \& {Ivezi{\'c}}}]{sheth03}
{Sheth}, R.~K., {Bernardi}, M., {Schechter}, P.~L., {et~al.} 2003, \apj, 594,
  225

\bibitem[{{Sullivan} {et~al.}(2000){Sullivan}, {Ellis}, {Nugent}, {Smail}, \&
  {Madau}}]{2000MNRAS.319..549S}
{Sullivan}, M., {Ellis}, R., {Nugent}, P., {Smail}, I., \& {Madau}, P. 2000,
  \mnras, 319, 549

\bibitem[{{Suyu} {et~al.}(2010){Suyu}, {Marshall}, {Auger}, {Hilbert},
  {Blandford}, {Koopmans}, {Fassnacht}, \& {Treu}}]{suyu10}
{Suyu}, S.~H., {Marshall}, P.~J., {Auger}, M.~W., {et~al.} 2010, \apj, 711, 201

\bibitem[{{Suyu} {et~al.}(2013){Suyu}, {Auger}, {Hilbert}, {Marshall}, {Tewes},
  {Treu}, {Fassnacht}, {Koopmans}, {Sluse}, {Blandford}, {Courbin}, \&
  {Meylan}}]{suyu13}
{Suyu}, S.~H., {Auger}, M.~W., {Hilbert}, S., {et~al.} 2013, \apj, 766, 70

\bibitem[{{Suyu} {et~al.}(2017){Suyu}, {Bonvin}, {Courbin}, {Fassnacht},
  {Rusu}, {Sluse}, {Treu}, {Wong}, {Auger}, {Ding}, {Hilbert}, {Marshall},
  {Rumbaugh}, {Sonnenfeld}, {Tewes}, {Tihhonova}, {Agnello}, {Blandford},
  {Chen}, {Collett}, {Koopmans}, {Liao}, {Meylan}, \& {Spiniello}}]{suyu17}
{Suyu}, S.~H., {Bonvin}, V., {Courbin}, F., {et~al.} 2017, \mnras, 468, 2590

\bibitem[{{Tewes} {et~al.}(2013){Tewes}, {Courbin}, {Meylan}, {Kochanek},
  {Eulaers}, {Cantale}, {Mosquera}, {Magain}, {Van Winckel}, {Sluse},
  {Cataldi}, {V{\"o}r{\"o}s}, \& {Dye}}]{tewes13}
{Tewes}, M., {Courbin}, F., {Meylan}, G., {et~al.} 2013, \aap, 556, A22

\bibitem[{{Tie} \& {Kochanek}(2017)}]{tk17}
{Tie}, S.~S., \& {Kochanek}, C.~S. 2017, ArXiv e-prints, arXiv:1707.01908

\bibitem[{{Treu} \& {Marshall}(2016)}]{cosmography}
{Treu}, T., \& {Marshall}, P.~J. 2016, \aapr, 24, 11

\bibitem[{{Vuissoz} {et~al.}(2008){Vuissoz}, {Courbin}, {Sluse}, {Meylan},
  {Chantry}, {Eulaers}, {Morgan}, {Eyler}, {Kochanek}, {Coles}, {Saha},
  {Magain}, \& {Falco}}]{vuissoz08}
{Vuissoz}, C., {Courbin}, F., {Sluse}, D., {et~al.} 2008, \aap, 488, 481

\bibitem[{{Wambsganss}(1990)}]{wthesis}
{Wambsganss}, J. 1990, PhD thesis, Thesis Ludwig-Maximilians-Univ., Munich
  (Germany, F.~R.).~Fakult{\"a}t f{\"u}r Physik., (1990)

\bibitem[{{Wambsganss}(1999)}]{mltree}
---. 1999, Journal of Computational and Applied Mathematics, 109, 353

\bibitem[{{Wang} \& {Wheeler}(2008)}]{specpol}
{Wang}, L., \& {Wheeler}, J.~C. 2008, \araa, 46, 433

\bibitem[{{Witt} \& {Mao}(1997)}]{witt97}
{Witt}, H.~J., \& {Mao}, S. 1997, \mnras, 291, 211

\bibitem[{{Wong} {et~al.}(2017){Wong}, {Suyu}, {Auger}, {Bonvin}, {Courbin},
  {Fassnacht}, {Halkola}, {Rusu}, {Sluse}, {Sonnenfeld}, {Treu}, {Collett},
  {Hilbert}, {Koopmans}, {Marshall}, \& {Rumbaugh}}]{wong17}
{Wong}, K.~C., {Suyu}, S.~H., {Auger}, M.~W., {et~al.} 2017, \mnras, 465, 4895

\bibitem[{{York} {et~al.}(2000){York}, {Adelman}, {Anderson}, {Anderson},
  {Annis}, {Bahcall}, {Bakken}, {Barkhouser}, {Bastian}, {Berman}, {Boroski},
  {Bracker}, {Briegel}, {Briggs}, {Brinkmann}, {Brunner}, {Burles}, {Carey},
  {Carr}, {Castander}, {Chen}, {Colestock}, {Connolly}, {Crocker}, {Csabai},
  {Czarapata}, {Davis}, {Doi}, {Dombeck}, {Eisenstein}, {Ellman}, {Elms},
  {Evans}, {Fan}, {Federwitz}, {Fiscelli}, {Friedman}, {Frieman}, {Fukugita},
  {Gillespie}, {Gunn}, {Gurbani}, {de Haas}, {Haldeman}, {Harris}, {Hayes},
  {Heckman}, {Hennessy}, {Hindsley}, {Holm}, {Holmgren}, {Huang}, {Hull},
  {Husby}, {Ichikawa}, {Ichikawa}, {Ivezi{\'c}}, {Kent}, {Kim}, {Kinney},
  {Klaene}, {Kleinman}, {Kleinman}, {Knapp}, {Korienek}, {Kron}, {Kunszt},
  {Lamb}, {Lee}, {Leger}, {Limmongkol}, {Lindenmeyer}, {Long}, {Loomis},
  {Loveday}, {Lucinio}, {Lupton}, {MacKinnon}, {Mannery}, {Mantsch}, {Margon},
  {McGehee}, {McKay}, {Meiksin}, {Merelli}, {Monet}, {Munn}, {Narayanan},
  {Nash}, {Neilsen}, {Neswold}, {Newberg}, {Nichol}, {Nicinski}, {Nonino},
  {Okada}, {Okamura}, {Ostriker}, {Owen}, {Pauls}, {Peoples}, {Peterson},
  {Petravick}, {Pier}, {Pope}, {Pordes}, {Prosapio}, {Rechenmacher}, {Quinn},
  {Richards}, {Richmond}, {Rivetta}, {Rockosi}, {Ruthmansdorfer}, {Sandford},
  {Schlegel}, {Schneider}, {Sekiguchi}, {Sergey}, {Shimasaku}, {Siegmund},
  {Smee}, {Smith}, {Snedden}, {Stone}, {Stoughton}, {Strauss}, {Stubbs},
  {SubbaRao}, {Szalay}, {Szapudi}, {Szokoly}, {Thakar}, {Tremonti}, {Tucker},
  {Uomoto}, {Vanden Berk}, {Vogeley}, {Waddell}, {Wang}, {Watanabe},
  {Weinberg}, {Yanny}, {Yasuda}, \& {SDSS Collaboration}}]{sdss}
{York}, D.~G., {Adelman}, J., {Anderson}, Jr., J.~E., {et~al.} 2000, \aj, 120,
  1579

\end{thebibliography}

\end{document}